\documentclass[a4paper,11pt]{article}
\pdfoutput=1 % if your are submitting a pdflatex (i.e. if you have
\usepackage{jheppub} % for details on the use of the package, please
\usepackage[T1]{fontenc} % if needed
\usepackage{slashed}
\usepackage{subcaption}
\usepackage{ucs}
\usepackage{hyperref}
\usepackage[utf8x]{inputenc}
\usepackage{amsmath}
\usepackage{amsfonts}
\usepackage{comment}
\usepackage{amssymb}
\usepackage{array}
\usepackage{epsfig}
\usepackage{slashed}
\title{\boldmath Momentum space spinning correlators and higher spin equations in three dimensions}

\author[a]{Sachin Jain,}
\author[b,c]{ Renjan Rajan John,}
\author[a]{Vinay Malvimat}

\affiliation[a]{Indian Institute of Science Education and Research, Homi Bhabha Rd, Pashan, Pune 411 008, India}
\affiliation[b]{Universit\`a del Piemonte Orientale, Dipartimento di Scienze e Innovazione Tecnologica, Viale T. Michel 11, I-15121 Alessandria, Italy}
\affiliation[c]{I.\,N.\,F.\,N. - sezione di Torino, Via P. Giuria 1, I-10125 Torino, Italy}

\emailAdd{sachin.jain@iiserpune.ac.in}
\emailAdd{vinaymm@iiserpune.ac.in}
\emailAdd{renjan.rajan@to.infn.it}
\abstract{

In this article, we explicitly compute in momentum space the three and four-point correlation functions involving scalar and spinning operators in the free bosonic and the free fermionic theory in three dimensions. We also evaluate the five-point function of the scalar operator in the free bosonic theory.
We discuss techniques which are more efficient  than the usual PV reduction to evaluate one loop integrals. Our techniques can be easily generalised to momentum space correlators of complicated spinning operators and to higher point functions. 
The three dimensional fermionic theory has the interesting feature  that the scalar operator ${\bar\psi} \psi$ is odd under parity.  To account for this, we develop a parity odd basis which is useful to write correlation functions involving spinning operators and an odd number of ${\bar\psi} \psi$ operators. We further study higher spin (HS) equations in momentum space which are algebraic in nature and hence simpler than their position space counterparts. We use them to solve for three-point functions involving  spinning operators without invoking conformal invariance. However, at the level of four-point functions, solving the HS equation requires additional constraints that come from conformal invariance and we could only verify that our explicit results solve the HS equation.
}

\begin{document} 
	\maketitle
	\raggedbottom
	%\pagebreak
	\flushbottom
	\section{Introduction}
	Conformal field theory (CFT) plays a central role in theoretical physics. It finds applicability in a host of phenomena that occur at very small length scales in particle physics, intermediate scales in condensed matter physics and very large scales in cosmology. With the development of the conformal bootstrap program, a lot of insight has been gained into the structure of CFT (see  \cite{Rattazzi:2008pe,Rychkov_2017,Poland_2019,simmonsduffin2016tasi} and references therein). Most of the development in the conformal bootstrap program is achieved in position space and its momentum space analogue has not gained much attention \cite{Corian__2013,Bzowski_2014,Isono_2018,Isono_2019,gillioz2019convergent}.  
	
Three-point correlation functions in momentum space have been determined through conformal invariance in  \cite{Corian__2013,Bzowski_2014,Bzowski:2015yxv,Bzowski_2016,Bzowski_2018,Corian__2018,Corian__2019,Bautista_2020}. The absence of the  analogue of conformal cross-ratios in momentum space makes the computation of four and higher point functions difficult. Furthermore, solving the special conformal Ward identity is in general hard as they are complicated second order differential equations in momentum variables. For specific higher point correlators limited progress has been made \cite{Gillioz:2018mto,Maglio:2019grh,corian2019fourpoint,Albayrak_2019,Albayrak:2018tam,Skvortsov:2018uru,Farrow_2019,Albayrak:2019asr,Albayrak:2020isk,gillioz2020scattering,Sleight:2019mgd,Sleight:2019hfp,Bzowski_2020,Serino:2020pyu}.  
Despite the difficulties it is desirable to understand CFT in momentum space. One of the reasons is its relation to Feynman graphs which are usually computed in momentum space. It is also significant for its applicability in the context of cosmology \cite{Mata_2013,Ghosh_2014,Kundu_2015,arkanihamed2015cosmological,arkanihamed2017cosmological,arkanihamed2018cosmological,Sleight:2019mgd,Sleight:2019hfp,baumann2019cosmological,baumann2020cosmological}. 
Although there are difficulties involved in imposing conformal invariance in momentum space there are some attractive features as well. For example, conformal blocks become simple products of two three-point functions. This is because descendants are related to primary operators by a simple multiplication of the momenta.

In the present article, we determine in momentum space  explicit three and four-point correlators of certain spinning operators in three dimensional free bosonic and free fermionic theories. In three dimensions, there exist parity odd correlators in the fermionic theory. We develop a basis to express parity odd three and four-point correlators. In order to compute scalar and spinning four-point correlators, we use some existing techniques and also develop some efficient methods to evaluate one loop three dimensional integrals. For work related to a detailed derivation of the operator algebra of conserved currents see \cite{Vasiliev} and references therein.

  The other aspect that we explore in this paper is higher-spin equations \cite{Giombi_2012,Aharony_2012,maldacena2011constraining,Maldacena_2013,Giombi_2017,Li:2019twz} in momentum space. In momentum space, these are algebraic, and hence simpler than their position space counterparts.  We utilise this aspect to solve for some of the three-point correlators involving spinning operators without invoking conformal invariance. However, solving the higher spin equation for  four-point correlators is more complicated. In this article we only verify that the explicitly computed four-point correlators solve the higher spin equations.

Our article is organized as follows. In Section \ref{Theory} we list the theories of interest to us and their corresponding action, spectrum and the explicit form of the higher spin operators. In Section \ref{sec3CWI} we describe the conformal Ward identities and the reconstruction formula for the three and four-point spinning correlators we study. In Section \ref{momentumbasis} we introduce the momentum basis for the transverse part of both parity odd and parity even higher spin correlators. In Section \ref{Resultsofexplicitcomputations} we provide explicit results for the three and four-point correlators of spinning and scalar operators in both free bosonic and free fermionic theories in three dimensions. In Section \ref{CorrelatorsfromhigherspinWardidentities} we illustrate the power of higher spin equations in solving for three-point spinning correlators without invoking conformal invariance. Furthermore, we verify that our explicit results for four-point functions solve the corresponding higher spin equations. In Appendix \ref{Comp} and \ref{Wick contractions}  we provide details of the computation of some of the correlators. In particular  we show how to utilise the Schouten identity and the inversion technique to evaluate integrals more efficiently than the usual PV reduction. In Appendix \ref{HSEQ4fermion} we describe some details of the higher spin equation. 
	 
	\section{List of Theories and Operators}\label{Theory}
		In this section we discuss the Lagrangian and the operator spectrum of theories of interest to us. 
	\subsection{Free Bosonic Theory}
	The simplest of CFTs we study in this article is the massless free bosonic ($FB$) theory in three dimensions. Let the scalar field $\phi$ be in the fundamental representation of $SU(N)$\footnote{Most of the discussion and the results for free bosons in this article are also applicable to the complex $U(1)$ boson.}. The action is given by :
	\begin{align}
S_{FB}=\int d^{3} x~ \partial^{\mu} \bar{\phi}~ \partial_{\mu} \phi \label{FBA}\,.
	\end{align}
	This theory exhibits a higher spin symmetry and hence has a tower of exactly conserved higher spin currents. We denote these operators as $J_s$ where $s$ represents the spin of the operator. In this article we deal with the scalar operator $J_0$ and the spin-one  and spin-two (stress tensor) conserved currents, $J_\mu$ and $T_{\mu\nu}$ respectively. In the free bosonic theory they are given by :
	\begin{align}
	J_{0}(x)&=\bar{\phi}\phi(x)\nonumber\\[5pt]
	J_{\mu}(x)&=\bar{\phi} \partial_{\mu}\phi(x)-\phi\partial_{\mu}\bar{\phi}(x)\nonumber\\[5pt]
	T_{\mu\nu}(x)&=\frac{3}{8}(\partial_{\mu} \bar{\phi}\partial_{\nu}\phi+\partial_{\nu} \bar{\phi}\partial_{\mu}\phi)-\frac{1}{8}( \bar{\phi}\partial_{\mu}\partial_{\nu}\phi+\phi\partial_{\mu}\partial_{\nu} \bar{\phi})-\frac{1}{4}g_{\mu\nu}\partial_{\rho} \bar{\phi} \partial^{\rho}\phi\notag\\&
	\hspace{.5cm}+\frac{g_{\mu\nu}}{24}(\phi\Box  \bar{\phi}+ \bar{\phi}\Box \phi)\,.\label{CCB}
	\end{align}
	We have chosen the stress tensor in such a way that it is explicitly traceless.
	In momentum space these operators take the following form :
	\begin{align}
	J_0( \boldsymbol{k})&=\int d^3l ~\bar{\phi}(l)\,\phi(k-l)\label{FBS0}\\
	J_\mu( \boldsymbol{k})&=\int d^3l ~(2 l-k)_\mu\,\bar{\phi}(l)\,\phi(k-l)\label{FBS1}\\
	T_{\mu\nu}( \boldsymbol{k})&=\int d^3l ~ H_{\mu \nu}(l,k)\,\bar{\phi}(l)\,\phi(k-l)\,,\label{defbstr}
	\end{align}
where
\begin{align}
	H_{\mu\nu}(l,k)&=-\frac{3}{8}(l_{\mu}(k-l)_{\nu}+(k-l)_{\mu}\,l_{\nu})+\frac{1}{8}((k-l)_{\mu}(k-l)_{\nu}+l_{\mu}\,l_{\nu})+\frac{1}{4}g_{\mu \nu}\,l\cdot (k-l)\notag\\&\hspace{.5cm}-\frac{g_{\mu\nu}}{24}(l^2+(k-l)^2)\,.
	\end{align}
	 These operators are part of the single-trace primary operators of the CFT and their scaling dimensions are given by :
	\begin{align}
	\Delta_{J_0}&=1\notag\\
\Delta_{J_s}&=s+1,~~~s\ge1\,.\label{SDB}
	\end{align}
	The scaling dimensions do not receive any anomalous corrections as the currents are exactly conserved. 
	\subsection{Free Fermionic theory}
	\label{FFTdetails}
	We will now discuss the theory of free massless fermions ($FF$) in three dimensions.
As in the free bosonic theory, we consider the fermionic field $\psi$ also to be in the fundamental representation of $SU(N)$. The action for this theory is as follows :
	\begin{align}
	S_{FF}=\int d^{3} x\, i\,\bar{\psi}\,\slashed{\partial} \psi \label{FFA}\,.
	\end{align}
	The operator spectrum of this CFT also has a  tower of exactly conserved currents. These are the single trace primaries of the CFT, one for each spin $s$. In the free fermionic theory, for $s=0$, $s=1$, and $s=2$ these are given by :
\begin{align}
J_0&=\bar{\psi}\,\psi(x)\notag\\
J_{\mu}(x)&=i\,\bar{\psi}\,\gamma_{\mu}\,\psi(x)\notag\\
T_{\mu\nu}(x)& =\frac{1}{4}\big[\bar{\psi}\,(\gamma_{\mu}\,\partial_{\nu}+\gamma_{\nu}\,\partial_{\mu})\psi-\partial_{\nu}\bar{\psi}\,\gamma_{\mu}\,\psi -\partial_{\mu}\bar{\psi}\,\gamma_{\nu}\,\psi\big]\,.\label{CCF}
\end{align}
In  momentum space these operators take the following form :
 \begin{align}
 J_0( \boldsymbol{k})&= \int d^3l ~\bar{\psi}(l)\,\psi(k-l)\label{FFS0}\\
 J_{\mu}( \boldsymbol{k})&= \int d^3l ~\bar{\psi}(l)\,\gamma_{\mu}\,\psi(k-l)\label{FFS1}\\
	T_{\mu\nu}( \boldsymbol{k})&=\frac{1}{4}\int d^3l ~~\bar{\psi}(l)\big[\gamma_{\mu}\,(2l-k)_{\nu} +\gamma_{\nu}\,(2l-k)_{\mu}\big]\psi(k-l)\,.\label{FFS2}
	\end{align}
The scaling dimensions of these operators are given by :
\begin{align}
\Delta_{J_0}&=2\notag\\
\Delta_{J_s}&=s+1~~~s\ge1\,.\label{SDF}
\end{align}
	Note that the spectrum of higher spin operators in the free fermionic theory and the free bosonic theory \eqref{SDB} are the same except for the scalar operators. The parity of the scalar operator in the two theories differ; even in the bosonic theory and odd in the fermionic theory.

\section{Conformal Ward identities}
\label{sec3CWI}
In this section we describe in detail the conformal Ward identities obeyed by correlation functions involving scalar and spinning operators \cite{Bzowski_2014}. We then focus on correlation functions involving conserved currents and implement the constraints coming from diffeomorphism invariance (conservation laws) and the Weyl Ward identity (trace Ward identity).
	
We consider the $n$-point function of primary operators $\mathcal O_1,\ldots,\mathcal O_n$ in a CFT and denote it by $\langle\mathcal O_1(\boldsymbol{k}_1)\mathcal O_2( \boldsymbol{k}_2)\ldots \mathcal O_n( \boldsymbol{k}_n)\rangle$. The operators can have a non-zero spin and hence have Lorentz indices, but we suppress them here for sake of brevity. We denote the correlator with the momentum conserving delta function stripped off using double brackets as $\langle\langle\,\mathcal O_1( \boldsymbol{k}_1)\,\mathcal O_2( \boldsymbol{k}_2)\,\ldots\,\mathcal O_n( \boldsymbol{k}_n)\,\rangle\rangle$ :
\begin{align}
\langle\,\mathcal O_1( \boldsymbol{k}_1)\,\ldots\,\mathcal O_n( \boldsymbol{k}_n)\,\rangle=(2\pi)^d\delta^{(3)}( \boldsymbol{k}_1+\ldots+ \boldsymbol{k}_n)\langle\langle\,\mathcal O_1( \boldsymbol{k}_1)\,\ldots\,\mathcal O_n( \boldsymbol{k}_n)\,\rangle\rangle\,.
\end{align}
We will now discuss the Ward identities that $\langle\langle\,\mathcal O_1( \boldsymbol{k}_1)\,\ldots\,\mathcal O_n( \boldsymbol{k}_n)\rangle\rangle$ satisfies. We denote the conformal dimension of $\mathcal O_i$ as $\Delta_i$. 
\subsection{Dilatation and Special conformal Ward identities}
The dilatation Ward identity on an $n$-point correlator with scalar and tensor insertions is given by  \cite{Bzowski_2014} :
\begin{align}
0=\left[-(n-1)d+\sum_{j=1}^n\,\Delta_j-\sum_{j=1}^{n-1}\,k_j^\alpha\,\frac{\partial}{\partial k_j^\alpha}\right]\langle\langle\,\mathcal O_1( \boldsymbol{k}_1)\,\ldots\,\mathcal O_n( \boldsymbol{k}_n)\,\rangle\rangle\,.
\end{align}
This imposes the following scaling behaviour on the correlator :
\begin{align}
\langle\langle\,\mathcal O_1(\lambda\,\boldsymbol{k}_1)\,\ldots\,\mathcal O_n(\lambda\,\boldsymbol{k}_n)\,\rangle\rangle&=\lambda^{-\left[(n-1)d-\sum_{i=1}^n\Delta_i\right]}\langle\langle\,\mathcal O_1(\boldsymbol{k}_1)\,\ldots\,\mathcal O_n(\boldsymbol{k}_n)\,\rangle\rangle\,.
\end{align} 
The special conformal Ward identity on the $n$-point correlator of scalar primaries is \cite{Bzowski_2014} :
\begin{align}
0=\sum_{j=1}^{n-1}\left[2(\Delta_j-d)\frac{\partial}{\partial k_{j}^{\kappa}}-2k_j^\alpha
\frac{\partial}{\partial k_{j}^{\alpha}}\frac{\partial}{\partial k_{j}^{\kappa}}+k_j^\kappa\frac{\partial}{\partial k_{j}^{\alpha}}\frac{\partial}{\partial k_{j\alpha}}\right]\langle\langle\,\mathcal O_1(\boldsymbol{k}_1)\,\ldots\,\mathcal O_n(\boldsymbol{k}_n)\,\rangle\rangle\,.
\end{align}
When the correlator involves spinning operators the special conformal Ward identity is modified by an additional differential operator that mixes the tensor indices of the correlator. The action of the additional differential operator on the correlator takes the following form  \cite{Bzowski_2014} :
\begin{align}
2\sum_{j=1}^{n-1}\,\sum_{k=1}^{n_j}\,\left(\delta^{\mu_{jk}\kappa}\,\frac{\partial}{\partial k_j^{\alpha_{jk}}}-\delta^\kappa_{\alpha_{jk}}\frac{\partial}{\partial k_{j_{\mu_{jk}}}}\right)\langle\langle\,\mathcal O_1^{\mu_{11}\ldots\mu_{1r_1}}( \boldsymbol{k}_1)\,\ldots \mathcal O_j^{\mu_{j1}\ldots\alpha_{jk}\ldots\mu_{jr_j}}( \boldsymbol{k}_j)\,\ldots \mathcal O_n^{\mu_{n1}\ldots\mu_{nr_n}}( \boldsymbol{k}_n)\,\rangle\rangle\,.
\end{align}
For example, in three dimensions the dilatation and special conformal Ward identities on the $\langle\langle J_\mu J_\nu J_0\rangle\rangle$ correlator when the scalar operator $J_0$ has dimension 1 take the form :
\begin{align}
0&=\left[1+k_1^\alpha\,\frac{\partial}{\partial k_1^\alpha}+k_2^\alpha\,\frac{\partial}{\partial k_2^\alpha}\right]\langle\langle\,J_\mu( \boldsymbol{k}_1)\,J_\nu( \boldsymbol{k}_2)\, J_0( \boldsymbol{k}_3)\,\rangle\rangle\cr
0&=\left[-2\frac{\partial}{\partial k_1^\kappa}-2\frac{\partial}{\partial k_2^\kappa}-2k_1^{\alpha}\,\frac{\partial}{\partial k_1^{\alpha}}\,\frac{\partial}{\partial k_1^\kappa}-2k_2^{\alpha}\frac{\partial}{\partial k_2^{\alpha}}\,\frac{\partial}{\partial k_2^\kappa}+k_{1\kappa}\,\frac{\partial}{\partial k_1^{\alpha}}\,\frac{\partial}{\partial k_{1,\alpha}}+k_{2\kappa}\,\frac{\partial}{\partial k_2^{\alpha}}\,\frac{\partial}{\partial k_{2,\alpha}}\right]\cr
&\hspace{3cm}\langle\langle\,J_\mu( \boldsymbol{k}_1)\,J_\nu( \boldsymbol{k}_2)\, J_0( \boldsymbol{k}_3)\,\rangle\rangle\cr
&\hspace{1cm}+2\left(\delta_{\mu\kappa}\,\frac{\partial}{\partial k_{1,\alpha_1}}-\delta_{\kappa}^{\alpha_1}\frac{\partial}{\partial k_{1}^{\mu}}\right)\langle\langle\,J_{\alpha_1}( \boldsymbol{k}_1)\,J_\nu( \boldsymbol{k}_2)\, J_0( \boldsymbol{k}_3)\,\rangle\rangle\cr
&\hspace{1cm}+2\left(\delta_{\nu\kappa}\,\frac{\partial}{\partial k_{2,\alpha_2}}-\delta_{\kappa}^{\alpha_2}\frac{\partial}{\partial k_{2}^{\nu}}\right)\langle\langle\,J_{\mu}( \boldsymbol{k}_1)\,J_{\alpha_2}( \boldsymbol{k}_2)\, J_0( \boldsymbol{k}_3)\,\rangle\rangle=0\,.
\end{align}
We will now discuss the Ward identities resulting from diffeomorphism invariance and Weyl invariance.
\subsection{Diffeomorphism and Weyl Ward identities: Local and transverse parts of the correlator}
For spinning correlators with conserved currents, the corresponding conservation laws lead to new Ward identities. Such identities allow us to express correlation functions as the sum of  a transverse part and a local part :
\begin{align}
\label{transversepluslocal}
&\langle\langle\,\mathcal O_1^{\mu_{11}\ldots\mu_{1r_1}}( \boldsymbol{k}_1)\,\ldots \mathcal O_j^{\mu_{j1}\ldots\alpha_{jk}\ldots\mu_{jr_j}}( \boldsymbol{k}_j)\,\ldots \mathcal O_n^{\mu_{n1}\ldots\mu_{nr_n}}( \boldsymbol{k}_n)\,\rangle\rangle\nonumber\\[5pt]
&=\langle\langle \mathcal O_1^{\mu_{11}\ldots\mu_{1r_1}}( \boldsymbol{k}_1)\,\ldots \mathcal O_j^{\mu_{j1}\ldots\alpha_{jk}\ldots\mu_{jr_j}}( \boldsymbol{k}_j)\,\ldots \mathcal O_n^{\mu_{n1}\ldots\mu_{nr_n}}( \boldsymbol{k}_n)\,\rangle\rangle_{\text{transverse}}\nonumber\\[5pt]
&\hspace{1cm}+\langle\langle\mathcal O_1^{\mu_{11}\ldots\mu_{1r_1}}( \boldsymbol{k}_1)\,\ldots \mathcal O_j^{\mu_{j1}\ldots\alpha_{jk}\ldots\mu_{jr_j}}( \boldsymbol{k}_j)\,\ldots \mathcal O_n^{\mu_{n1}\ldots\mu_{nr_n}}( \boldsymbol{k}_n)\,\rangle\rangle_{\text{local}}\,.
\end{align}
Here the transverse part is such that, $\forall a=1,\ldots, n$ and $\forall i=1,\ldots,r_a$
\begin{align}
&k_{a,\mu_{ai}}\langle\langle \mathcal O_1^{\mu_{11}\ldots\mu_{1r_1}}( \boldsymbol{k}_1)\,\ldots \mathcal O_j^{\mu_{j1}\ldots\alpha_{jk}\ldots\mu_{jr_j}}( \boldsymbol{k}_j)\,\ldots \mathcal O_n^{\mu_{n1}\ldots\mu_{nr_n}}( \boldsymbol{k}_n)\,\rangle\rangle_{\text{transverse}}=0\,.
\end{align}
The transverse part may then be written in a suitable basis of projectors which makes the transversality explicit. The local part is completely determined in terms of lower point correlators. In this work we will mostly be interested in correlators with a single insertion of the stress-tensor or two insertions of spin-one currents with scalar primaries. We will now discuss the transverse Ward identity obeyed by such correlators.
\subsubsection{Diffeomorphism/Transverse Ward identity} 
We first discuss the transverse Ward identity obeyed by a correlator with a single stress-tensor insertion.
\\[5pt]
$\mathbf{\langle T_{\mu\nu}\,J_0\ldots J_0\rangle}$
\\[5pt]
For a three-point correlator with a stress-tensor insertion  the transverse Ward identity takes the following form :
\begin{align}
\label{TWITJ0J0}
k_1^\mu\langle\langle T_{\mu\nu}( \boldsymbol{k}_1) J_0( \boldsymbol{k}_2) J_0( \boldsymbol{k}_3)\rangle\rangle &=-k_{2\nu}\langle J_0( \boldsymbol{k}_3)J_0(- \boldsymbol{k}_3)\rangle
-k_{3\nu} \langle J_0( \boldsymbol{k}_2)J_0(- \boldsymbol{k}_2)\rangle\nonumber\\[5pt]
k_1^\nu\langle\langle T_{\mu\nu}( \boldsymbol{k}_1) J_0( \boldsymbol{k}_2) J_0( \boldsymbol{k}_3)\rangle\rangle&=-k_{2\mu}\langle J_0( \boldsymbol{k}_3)J_0(- \boldsymbol{k}_3)\rangle
-k_{3\mu} \langle J_0( \boldsymbol{k}_2)J_0(- \boldsymbol{k}_2)\rangle\,.
\end{align}
For the four-point correlator it is :
\begin{align}
k_{1\mu} \langle\langle T^{\mu\nu}( \boldsymbol{k}_1) J_0( \boldsymbol{k}_2) J_0( \boldsymbol{k}_3)J_0( \boldsymbol{k}_4)\rangle\rangle&=-k_2^{\nu}\langle\langle J_0( \boldsymbol{k}_1+ \boldsymbol{k}_2)J_0( \boldsymbol{k}_3)J_0( \boldsymbol{k}_4)\rangle\rangle\nonumber\\[5pt]
&\hspace{-3.5cm}-k_3^{\nu}\langle\langle J_0( \boldsymbol{k}_1+ \boldsymbol{k}_3)J_0( \boldsymbol{k}_2)J_0( \boldsymbol{k}_4)\rangle\rangle-k_4^{\nu}\langle\langle J_0( \boldsymbol{k}_1+ \boldsymbol{k}_4)J_0( \boldsymbol{k}_2)J_0( \boldsymbol{k}_3)\rangle\rangle\,.
%\end{align}
%and
%\begin{align}
\end{align}
and a similar identity upon the action of $k_{1\nu}$. 
%\begin{align}
%k_{1\nu}\langle\langle T^{\mu\nu}( \boldsymbol{k}_1) J_0( \boldsymbol{k}_2) J_0( \boldsymbol{k}_3)J_0( \boldsymbol{k}_4)\rangle\rangle&=-k_2^{\mu}\langle\langle J_0( \boldsymbol{k}_3+ \boldsymbol{k}_4)J_0( \boldsymbol{k}_3)J_0( \boldsymbol{k}_4)\rangle\rangle\nonumber\\[5pt]
%&\hspace{-2.5cm}-k_3^{\mu}\langle\langle J_0( \boldsymbol{k}_2+ \boldsymbol{k}_4)J_0( \boldsymbol{k}_2)J_0( \boldsymbol{k}_4)\rangle\rangle-k_4^{\mu}\langle\langle J_0( \boldsymbol{k}_2+ \boldsymbol{k}_3)J_0( \boldsymbol{k}_2)J_0( \boldsymbol{k}_3)\rangle\rangle
%\end{align}
%
This can be easily generalised to the $n$-point case ($n\ge 4)$ :
\begin{align}\label{trTwrd}
\hspace{-.3cm}k_1^\mu\langle\langle\,T_{\mu\nu}( \boldsymbol{k}_1)\,J_0( \boldsymbol{k}_2)\,\ldots\,J_0( \boldsymbol{k}_n)\rangle\rangle=-\sum_{i=2}^{n}\,k_{i\nu}\big\langle\big\langle\,J_0\big(\textstyle\sum_{\ell\ne i,\,\ell=2}^n\,( \boldsymbol{k}_\ell\big)\textstyle\prod_{j\ne i,\,j=2}^{n}\,J_0(\boldsymbol{k}_j)\big\rangle\big\rangle\,.
\end{align}
%Let us note that, for the bosonic theory we have chosen to work with a stress tensor which is explicitly traceless \eqref{defbstr}. Thus the Ward identity takes a slightly different form than \eqref{trTwrd}, as will be discussed below. 
We will now consider correlators with the insertion of two spin-one currents.
\\[5pt]
$\mathbf{\langle J_{\mu}\,J_{\nu}\,J_0\ldots J_0\rangle}$
\\[5pt]
Let $A_\mu$ be the potential that sources the current $J_\mu$.
The transversality condition for the three-point correlator with two spin-one currents is given by \cite{Bzowski_2014} :
\begin{align}
\label{transversalityJmuJnu}
k_1^\mu\langle\langle J_{\mu}( \boldsymbol{k}_1) J_\nu( \boldsymbol{k}_2) J_0(\boldsymbol{k}_3)\rangle\rangle=&k_{1}^{\mu}\left\langle\left\langle \frac{\partial J_\mu}{\partial A^\nu}( \boldsymbol{k}_1, \boldsymbol{k}_2)\,J_0(\boldsymbol{k}_3)\right\rangle\right\rangle\,. 
\end{align}
For the three-point correlator in the bosonic theory this gives :
\begin{align}
\label{transverseWIfreeboson}
k_1^\mu\langle\langle J_{\mu}( \boldsymbol{k}_1) J_\nu( \boldsymbol{k}_2) J_0(\boldsymbol{k}_3)\rangle\rangle=&k_{1\nu}\langle J_0( \boldsymbol{k}_3)J_0(- \boldsymbol{k}_3)\rangle\nonumber\\[5pt]
k_2^\nu\langle\langle J_{\mu}( \boldsymbol{k}_1) J_\nu( \boldsymbol{k}_2) J_0(\boldsymbol{k}_3)\rangle\rangle=&k_{2\nu}\langle J_0( \boldsymbol{k}_3)J_0(- \boldsymbol{k}_3)\rangle\,. 
\end{align}
This can be readily extended to the four-point case as follows :
\begin{align}
\label{transverseWIJmuJnuJoJo}
k_1^{\mu}\langle\langle J_{\mu}( \boldsymbol{k}_1) J_\nu( \boldsymbol{k}_2) J_0(\boldsymbol{k}_3)J_0(\boldsymbol{k}_4)\rangle\rangle=k_{1\nu}\langle\langle J_0( \boldsymbol{k}_1+ \boldsymbol{k}_2)J_0( \boldsymbol{k}_3)J_0( \boldsymbol{k}_4)\rangle\rangle\,.
%k_2^{\nu}\langle\langle J_{\mu}( \boldsymbol{k}_1) J_\nu( \boldsymbol{k}_2) J_0(\boldsymbol{k}_3J_0(\boldsymbol{k}_4)\rangle\rangle=k_{2\mu}\langle J_0( \boldsymbol{k}_3+ \boldsymbol{k}_4)J_0( \boldsymbol{k}_3)J_0( \boldsymbol{k}_4)\rangle
\end{align}
A similar identity holds for the action of $k_2^{\nu}$ on the correlator.
For general $n$-point functions ($n\ge 4)$ with two spin-one currents, the transverse Ward identity takes the form :
\begin{align}
k_1^\mu\langle\langle\,J_{\mu}( \boldsymbol{k}_1)\,J_{\nu}( \boldsymbol{k}_2)\,J_0( \boldsymbol{k}_3)\,\ldots\,J_0( \boldsymbol{k}_n)\rangle\rangle=k_{1\nu}\big\langle\big\langle\,J_0\big(\textstyle\sum_{i=3}^n\, \boldsymbol{k}_i\big)\textstyle\prod_{j=3}^{n}\,J_0( \boldsymbol{k}_j)\big\rangle\big\rangle\,.
%k_2^\mu\langle\langle\,J_{\mu}( \boldsymbol{k}_1)\,J_{\nu}( \boldsymbol{k}_2)\,J_0( \boldsymbol{k}_3)\,\ldots\,J_0( \boldsymbol{k}_n)\rangle\rangle=k_{2\nu}\big\langle\,J_0\big(\textstyle\sum_{i=3}^n\, \boldsymbol{k}_i\big)\textstyle\prod_{j=3}^{n}\,J_0( \boldsymbol{k}_j)\big\rangle
\end{align}
For the fermionic theory, since the current takes the simple form $\bar\psi\gamma^\mu\psi$ the transverse Ward identity is trivial and gives :
\begin{align}
\label{transverseWIfermion}
k_1^\mu\langle\langle\,J_{\mu}( \boldsymbol{k}_1)\,J_{\nu}( \boldsymbol{k}_2)\,J_0( \boldsymbol{k}_3)\,\ldots\,J_0( \boldsymbol{k}_n)\rangle\rangle=0\,.
\end{align}
We will now discuss the Ward identities associated to Weyl invariance.
\subsubsection{Weyl/Trace Ward identities}
The trace Ward identity for a correlator involving stress-tensor and scalars with dimension $\Delta$ is given by :
\begin{align}
\label{traceWI3pt}
\delta_{\mu\nu}\langle\langle T^{\mu\nu}( \boldsymbol{k}_1)\,J_0(\boldsymbol{k}_2)\,J_0( \boldsymbol{k}_3)\rangle\rangle&=(d-\Delta)\big[\langle J_0( \boldsymbol{k}_2)J_0(- \boldsymbol{k}_2)\rangle+\langle J_0( \boldsymbol{k}_3)J_0(- \boldsymbol{k}_3)\rangle\big]\,.
\end{align}
For the four-point correlator this becomes :
\begin{align}
\label{traceWI4pt}
\delta_{\mu\nu}\langle\langle T^{\mu\nu}( \boldsymbol{k}_1)\,J_0( \boldsymbol{k}_2)\,J_0(\boldsymbol{k}_3)\,J_0(\boldsymbol{k}_4)\rangle\rangle&=(d-\Delta)\big[\langle\langle J_0( \boldsymbol{k}_1+ \boldsymbol{k}_2)J_0( \boldsymbol{k}_3)J_0( \boldsymbol{k}_4)\rangle\rangle\nonumber\\[5pt]
&\hspace{-3cm}+\langle\langle J_0( \boldsymbol{k}_1+ \boldsymbol{k}_3)J_0( \boldsymbol{k}_2)J_0( \boldsymbol{k}_4)\rangle\rangle+\langle\langle J_0( \boldsymbol{k}_1+ \boldsymbol{k}_4)J_0( \boldsymbol{k}_2)J_0( \boldsymbol{k}_3)\rangle\rangle\big]\,.
\end{align}
\subsection{ Reconstruction formula}
\label{reconstructionformuladetails}
As noted earlier, correlation functions involving conserved currents can be written as the sum of a transverse part and a local part. The local part of these correlators is fully determined in terms of lower point functions. In this sub-section, we give explicit expressions for the local part of correlators involving either a stress-tensor insertion or two spin-one insertions and scalars of dimension $\Delta$ in $d$ dimensions  \cite{corian2019fourpoint}. 
\\[5pt]
$\mathbf{\langle T_{\mu\nu}\,J_0\ldots J_0\rangle}$
\\[5pt]
For the three-point correlator with a stress-tensor insertion we have :
\begin{align}
\label{Tj0j0reconstruction}
\hspace{-.5cm}\langle\langle T_{\mu\nu}( \boldsymbol{k}_1)\,J_0( \boldsymbol{k}_2)\,J_0( \boldsymbol{k}_3)\rangle\rangle=\langle\langle T_{\mu\nu}( \boldsymbol{k}_1)\,J_0( \boldsymbol{k}_2)\,J_0( \boldsymbol{k}_3)\rangle\rangle_{\text{transverse}}+
\langle\langle T_{\mu\nu}( \boldsymbol{k}_1)\,J_0( \boldsymbol{k}_2)\,J_0( \boldsymbol{k}_3)\rangle\rangle_{\text{local}}
\end{align}
where the local part is given by \cite{corian2019fourpoint} :
\begin{align}\label{localtermTJ0J0} 
\langle\langle &T_{\mu\nu}( \boldsymbol{k}_1) J_0( \boldsymbol{k}_2) J_0( \boldsymbol{k}_3)\rangle\rangle_{\text{local}}=\frac{2}{k_1^2}\bigg[-k_{(1\mu}k_{2\nu)}\langle J_0( \boldsymbol{k}_3)J_0(- \boldsymbol{k}_3)\rangle-k_{(1\mu}k_{3\nu)}\langle J_0( \boldsymbol{k}_2)J_0(- \boldsymbol{k}_2)\rangle\bigg]\cr
&\hspace{4cm}+\frac{d-\Delta}{d-1}\left(g_{\mu\nu}-\frac{k_{1\mu}k_{1\nu}}{k_1^2}\right)\bigg[\langle J_0( \boldsymbol{k}_2)J_0(- \boldsymbol{k}_2)\rangle+\langle J_0( \boldsymbol{k}_3)J_0(- \boldsymbol{k}_3)\rangle\bigg]\notag\\&\hspace{.8cm}
-\frac{1}{d-1}\bigg(g_{\mu\nu}+(d-2)\frac{k_{1\mu}k_{1\nu}}{k_1^2}\bigg)\bigg[-\frac{k_1\cdot k_2}{k_1^2}\langle J_0( \boldsymbol{k}_3)J_0(- \boldsymbol{k}_3)\rangle - \frac{k_1\cdot k_3}{k_1^2} \langle J_0( \boldsymbol{k}_2)J_0(- \boldsymbol{k}_2)\rangle\bigg]\,.
\end{align}
For the four-point correlator the local part takes the form \cite{corian2019fourpoint} :
\begin{align}
\label{localtermTJ0J0J0}
\langle\langle& T_{\mu\nu}( \boldsymbol{k}_1) J_0( \boldsymbol{k}_2) J_0( \boldsymbol{k}_3)J_0( \boldsymbol{k}_4)\rangle\rangle_{\text{local}}=\frac{2}{k_1^2}\bigg[-k_{(1\mu}k_{2\nu)}\langle\langle J_0( \boldsymbol{k}_1+ \boldsymbol{k}_2)J_0( \boldsymbol{k}_3)J_0( \boldsymbol{k}_4)\rangle\rangle\notag\\&-k_{(1\mu}k_{3\nu)}\langle\langle J_0( \boldsymbol{k}_1+ \boldsymbol{k}_3)J_0( \boldsymbol{k}_2)J_0( \boldsymbol{k}_4)\rangle\rangle-k_{(1\mu}k_{4\nu)}\langle\langle J_0( \boldsymbol{k}_1+ \boldsymbol{k}_4)J_0( \boldsymbol{k}_2)J_0( \boldsymbol{k}_3)\rangle\rangle\bigg]\notag\\&+\frac{d-\Delta}{d-1}\left(g_{\mu\nu}-\frac{k_{1\mu}k_{1\nu}}{k_1^2}\right)\bigg[\langle\langle J_0( \boldsymbol{k}_1+ \boldsymbol{k}_2)J_0( \boldsymbol{k}_3)J_0( \boldsymbol{k}_4)\rangle\rangle+\langle\langle J_0( \boldsymbol{k}_1+ \boldsymbol{k}_3)J_0( \boldsymbol{k}_2)J_0( \boldsymbol{k}_4)\rangle\rangle\notag\\&\hspace{5cm}
+\langle\langle J_0( \boldsymbol{k}_1+ \boldsymbol{k}_4)J_0( \boldsymbol{k}_2)J_0( \boldsymbol{k}_3)\rangle\rangle\bigg]\notag\\&
-\frac{1}{(d-1)}\bigg(g_{\mu\nu}+(d-2)\frac{k_{1\mu}k_{1\nu}}{k_1^2}\bigg)\bigg[-\frac{k_1\cdot k_2}{k_1^2}\langle\langle J_0( \boldsymbol{k}_1+ \boldsymbol{k}_2)J_0( \boldsymbol{k}_3)J_0( \boldsymbol{k}_4)\rangle\rangle\notag\\&-\frac{k_1\cdot k_3}{k_1^2}\langle\langle J_0( \boldsymbol{k}_1+ \boldsymbol{k}_3)J_0( \boldsymbol{k}_2)J_0( \boldsymbol{k}_4)\rangle\rangle- \frac{k_1\cdot k_4}{k_1^2}\langle\langle J_0( \boldsymbol{k}_1+ \boldsymbol{k}_4)J_0( \boldsymbol{k}_2)J_0( \boldsymbol{k}_3)\rangle\rangle\bigg]\,.
\end{align}
Note that the symmetrization bracket comes with a normalisation of 1/2.
\\[5pt]
$\mathbf{\langle J_{\mu}J_{\nu}\,J_0\ldots J_0\rangle}$
\\[5pt]
For the three-point correlator with two spin-one insertions one can do a similar splitting as in \eqref{Tj0j0reconstruction} :
\begin{align}
\label{jmujnu3pful}
\langle\langle J_{\mu}( \boldsymbol{k}_1)\,J_{\nu}( \boldsymbol{k}_2)\,J_0( \boldsymbol{k}_3)\rangle\rangle&=\langle\langle J_{\mu}( \boldsymbol{k}_1)\,J_\nu( \boldsymbol{k}_2)\,J_0( \boldsymbol{k}_3)\rangle\rangle_{\text{transverse}}+\langle\langle J_{\mu}( \boldsymbol{k}_1)\,J_\nu( \boldsymbol{k}_2)\,J_0( \boldsymbol{k}_3)\rangle\rangle_{\text{local}}\,.
\end{align}
In the bosonic theory the local term takes the form :
\begin{align}
\label{jmujnuj0local}
\hspace{-.5cm}\langle\langle J_{\mu}( \boldsymbol{k}_1) J_\nu( \boldsymbol{k}_2) J_0( \boldsymbol{k}_3)\rangle\rangle_{\text{local}}=&\langle J_0( \boldsymbol{k}_3)J_0(- \boldsymbol{k}_3)\rangle\left(\frac{k_{1\mu}k_{1\nu}}{k_1^2}+\frac{k_{2\mu}k_{2\nu}}{k_2^2}-\frac{k_{1\mu}k_{2\nu}\, k_1\cdot k_2}{k_1^2k_2^2}\right)\,.
\end{align}
For the four-point correlator in the bosonic theory the local term is given by :
\begin{align}\label{j14pt}
\langle\langle & J_{\mu}( \boldsymbol{k}_1) J_\nu( \boldsymbol{k}_2) J_0(\boldsymbol{k}_3)J_0( \boldsymbol{k}_4)\rangle\rangle_{\text{local}}=\frac{k_{1\mu}k_{1\nu}}{k_1^2}\langle\langle J_0( \boldsymbol{k}_1+ \boldsymbol{k}_2)J_0( \boldsymbol{k}_3)J_0( \boldsymbol{k}_4)\rangle\rangle\notag\\&+\frac{k_{2\mu}k_{2\nu}}{k_2^2}\langle\langle J_0( \boldsymbol{k}_1+ \boldsymbol{k}_2)J_0( \boldsymbol{k}_3)J_0( \boldsymbol{k}_4)\rangle\rangle -\frac{k_{1\mu}k_{2\nu}\,{k}_1\cdot {k}_2}{k_1^2k_2^2}\langle\langle J_0( \boldsymbol{k}_1+ \boldsymbol{k}_2)J_0( \boldsymbol{k}_3)J_0( \boldsymbol{k}_4)\rangle\rangle\,.
\end{align}
In the fermionic theory owing to the fact that the transverse Ward identity is trivial \eqref{transverseWIfermion} there is local part to correlators involving two spin-one currents.

We will now turn our attention to the transverse part of the correlators. To express the transverse part compactly, we need to work with a suitable momentum basis. We discuss this in the following section.
	\section{Momentum basis for the transverse part}
	\label{momentumbasis}
	 It is convenient to introduce a suitable basis to express the transverse part of correlators involving spinning operators. In this section we describe projector bases in momentum space. For parity even correlators, this has been discussed in \cite{Bzowski_2014}. We introduce another basis for such correlators, which is particularly useful in three dimensions. We also develop a basis to express parity odd correlators.
	
\subsection{Projector basis for  parity even correlators}
The simplest projector which projects orthogonal to the momentum $\boldsymbol{p}$ is :
	 \begin{align}
	 \pi_{\alpha}^{\mu}(\boldsymbol{p})&\equiv\delta_{\alpha}^{\mu}-\frac{p^{\mu}\,p_{\alpha}}{p^{2}}\,.\label{projk1}
	 \end{align}
	 It can be easily seen that this projector is orthogonal to $\boldsymbol{p}$ :
	 \begin{align}
	p_{\mu}\,\pi_{\alpha}^{\mu}(\boldsymbol{p})=p_{\alpha}\,\pi_{\alpha}^{\mu}(\boldsymbol{p})=0\,.
	 	\end{align}
	 	 For a correlator involving the stress tensor, one can use following projector :
		\begin{align}
	\Pi_{\alpha \beta}^{\mu \nu}(\boldsymbol{p})&\equiv\frac{1}{2}\left(\pi_{\alpha}^{\mu}(\boldsymbol{p})\,\pi_{\beta}^{\nu}(\boldsymbol{p})+\pi_{\beta}^{\mu}(\boldsymbol{p})\,\pi_{\alpha}^{\nu}(\boldsymbol{p})\right)-\frac{1}{2} \pi^{\mu \nu}(\boldsymbol{p})\,\pi_{\alpha \beta}(\boldsymbol{p})\,.\label{projk1k1}
	\end{align}
	  We can easily check that the projector is orthogonal to ${\boldsymbol{p}}$ and is traceless :
	 \begin{align}
p_{\mu}\,\Pi_{\alpha \beta}^{\mu \nu}(\boldsymbol{p})&=p_{\nu}\,\Pi_{\alpha \beta}^{\mu \nu}(\boldsymbol{p})=0\notag\\
g_{\mu\nu}\,\Pi_{\alpha \beta}^{\mu \nu}(\boldsymbol{p})&=0.
	 \end{align}
	The projector basis that appears in our computation of the correlator with two spin-one conserved currents  is as follows :
	\begin{align}
	\zeta_{\alpha \beta}^{\mu \nu}( \boldsymbol{p}_1, \boldsymbol{p}_2)\equiv\pi_{\alpha}^{\mu}\left( \boldsymbol{p}_1\right) \pi_{\beta}^{\nu}\left( \boldsymbol{p}_2\right)\,,\label{Projp1p2}
	\end{align}
	where $\pi_{\alpha}^{\mu}\left( \boldsymbol{p}_1\right)$ and $\pi_{\beta}^{\nu}\left( \boldsymbol{p}_2\right)$ can be obtained from \eqref{projk1}. Note that the above projector is orthogonal to   $\boldsymbol{p}_1$ in the $\mu$ index  and  orthogonal to $\boldsymbol{p}_2$  in the $\nu$ index :
		\begin{align}
p_{1\mu}\,\zeta_{\alpha \beta}^{\mu \nu}( \boldsymbol{p}_1, \boldsymbol{p}_2)=p_{2\nu}\,\zeta_{\alpha \beta}^{\mu \nu}( \boldsymbol{p}_1, \boldsymbol{p}_2)=0\,.
	\end{align}
\subsubsection{A basis involving the Levi-Civita tensor}

Quite interestingly, in three dimensions there exists a set of projectors which are expressed in terms of the Levi-Civita tensor and are equivalent to the ones discussed above. We notice that the projector in \eqref{projk1} can be expressed in terms of the Levi-Civita tensor as follows :
\begin{align}
\frac{\epsilon^{\mu k_1 \alpha}\,\epsilon_{\nu k_1 \alpha}}{k_1^2}=\pi^{\mu}_{\nu}(\boldsymbol{k}_1)\,.
\end{align}
Here we introduced the following notation for compactness :
\begin{align}
\label{enot}
\epsilon^{\mu k_1 \lambda}\equiv\epsilon^{\mu \delta \lambda}k_{1,\delta}\,.
\end{align}
An orthogonal and traceless projector equivalent to the one in \eqref{projk1k1} can be written as :
	\begin{align}
		\chi_{\alpha \beta}^{\mu \nu}(\boldsymbol{p})\equiv\frac{1}{2}\bigg(\epsilon^{\mu}_{ p \alpha}\,\epsilon^{\nu}_{ p \beta}+\epsilon^{\mu}_{ p \beta}\,\epsilon^{\nu}_{ p \alpha}\bigg)-\frac{1}{ p^2}\epsilon^{\mu\,p}_{~~~\lambda}\,\epsilon^{\nu\,p\, \lambda}\,\epsilon_{\alpha\,p\,\delta}\,\epsilon_{\beta\,p }^{~~~ \delta}\,.\label{Ek1k1}
	\end{align}
After some algebra one can show that :
\begin{equation}\label{rlation1}
	\chi_{\alpha \beta}^{\mu \nu}(\boldsymbol{p})=-p^2~ \Pi_{\alpha\beta}^{\mu\nu}(p)\,.
\end{equation}
The projector equivalent to the one in \eqref{Projp1p2} which is orthogonal to  $\boldsymbol{p}_1$ in the $\mu$ index  and  orthogonal to $\boldsymbol{p}_2$ in the $\nu$ index is expressed in terms of Levi-Civita tensor  as follows :
	\begin{align}\label{Ek1k2}
	\Sigma_{\alpha \beta}^{\mu \nu}( \boldsymbol{p}_1, \boldsymbol{p}_2)\equiv\epsilon^{\mu}_{ p_1\alpha}\,\epsilon^{\nu}_{ p_2\beta}\,.
	\end{align}
	This completes our discussion on parity even projectors. We will now describe a basis for parity odd correlators.
		\subsection{Projector basis for parity odd correlators}
		In three dimensions, apart from parity even correlators one has to deal with parity odd correlators as well. Hence it is useful to have a projector basis for such correlators. To obtain a parity odd basis one can use an odd number of Levi-Civita tensors. For example, the analogue of \eqref{projk1k1} and \eqref{Ek1k1} takes the form :
			\begin{align}
	\Delta_{\alpha \beta}^{\mu \nu}(\boldsymbol{p})\equiv\epsilon^{\mu p}_{\alpha}\,\pi^{\nu}_{\beta}(p)+\epsilon^{\nu p}_{\alpha}\,\pi^{\mu}_{\beta}(p)+\epsilon^{\mu p}_{\beta}\,\pi^{\nu}_{\alpha}(p)+\epsilon^{\nu p}_{\beta}\,\pi^{\mu}_{\alpha}(p)\,.\label{Ek1k1od}
		\end{align}
		Note that with this definition $\Delta_{\alpha \beta}^{\mu \nu}(\boldsymbol{p})$ is automatically orthogonal and traceless : 
		\begin{equation}
		\Delta_{\alpha \beta}^{\mu \nu}(\boldsymbol{p}) p_{\mu}=\Delta_{\alpha \beta}^{\mu \nu}(\boldsymbol{p}) p_{\nu}=\Delta_{\alpha \beta}^{\mu \nu}(\boldsymbol{p}) p^{\alpha}=\Delta_{\alpha \beta}^{\mu \nu}(\boldsymbol{p}) p^{\beta}=0\,,
		\end{equation} and
		\begin{equation}
		\Delta_{\alpha \beta}^{\mu \nu}(\boldsymbol{p}) g_{\mu\nu}=\Delta_{\alpha \beta}^{\mu \nu}(\boldsymbol{p}) g^{\alpha\beta}=0\,.
		\end{equation} 
Another useful basis is :
\begin{equation}
\Omega^{\mu \nu}(\boldsymbol{p_1,p_2})\equiv\epsilon^{\mu p_1}_{\alpha} \epsilon^{\nu p_2}_{\beta}\epsilon^{\alpha \beta p_i}\label{Ek1k2od1}\,,
\end{equation}
where $p_i$ on the R.H.S can be either $p_1$ or $p_2$. This basis is useful in describing the three-point function of two spin-one currents $J_{\mu}$ and the parity odd operator $J_0$ in the fermionic theory.
    
In some situations, one can define a parity odd projector\footnote{Although these are explicitly orthogonal to their momentum arguments, squaring them does not give back the same quantity. This can be easily seen by noting that squaring a parity odd quantity gives a parity even quantity. In this sense, these are different from usual projectors.} by simply multiplying the parity even projector with a Levi-Civita tensor. 
For example, in the case of four-point functions one can define a parity odd projector tensor as :
\begin{equation}\label{podproj}
{\rm{Proj}_{odd}}=\epsilon^{p_1\,p_2\,p_3} ~{\rm{Proj}_{even}}\,,
\end{equation}
where $\epsilon^{p_1\,p_2\,p_3}$ is defined following \eqref{enot}. However, it is important to note that  \eqref{podproj} may not always be very convenient. For example, to describe the parity odd correlator $\langle T_{\mu\nu}(p_1)J_0(p_2) J_0(p_3) J_0(p_4) \rangle$ in free fermionic theory, it is more convenient to use \eqref{Ek1k1od} than $\epsilon^{p_1\,p_2\,p_3}\,\Pi^{\mu \nu}_{\alpha\beta}(p_1)$.

  %  \section{Correlator definition} Basis both parity even and parity odd for three and four-point function+ Ward identities + conformal Ward identities + symmetry of coefficients.

%	\section{Higher-spin equation}
%	\subsection{Three-point function}
%	1. Free theory.
%	2. No of structure i  interractiong theory.
%	\subsection{Four-point function}
%	1. Free theory: Working in s,t,u variable.
%	2. No of structure in interacting theory. 
%	Correlators we are interested in and parity even and parity odd basis. Correlator in CS matter theory 
	 \section{Results of explicit computations} 
	 \label{Resultsofexplicitcomputations}
	 In this section we present explicit results for  correlation functions of interest to us in the free bosonic and free fermionic theories in three dimensions. 
\subsection{Free Bosonic  theory in $d=3$}
In this section we give the results for two, three and four-point functions involving operators of interest to us in the free bosonic theory.
\subsubsection{Two and Three-point functions}
The two-point function of the scalar operator is given by :
\begin{align}
\langle J_0( \boldsymbol{k})J_0(-\boldsymbol{k})\rangle=\frac{1}{8k}\,.
\end{align}
%\\[5pt]
$\mathbf{\langle \boldsymbol{J_{0}\,J_0\,J_0}\rangle}$
\\[5pt]
The three-point function of the scalar operator is given by :
\begin{align}
\label{j0j0j03pt}
\langle \langle J_0( \boldsymbol{k}_1)J_0( \boldsymbol{k}_2)J_0( \boldsymbol{k}_3)\rangle\rangle=\frac{1}{4\,k_1\,k_2\,k_3}\,.
\end{align}
%\\[5pt]
$\mathbf{ \boldsymbol\langle{J_{\mu}\,J_0\,J_0}\rangle}$
\\[5pt]
Let us now compute the simplest  three-point function with a spinning operator which is $\langle J_\mu J_0J_0\rangle$ where $J_{\mu}$ is the conserved spin-one current in the bosonic theory given by \eqref{FBS1}. Explicit Wick contractions show that this three-point function is zero :
\begin{align}
\langle\langle J_{\mu}( \boldsymbol{k}_1) J_0( \boldsymbol{k}_2) J_0( \boldsymbol{k}_3)\rangle\rangle=0\,.
\end{align}
This can be understood from the fact that $J_\mu$ and thereby the correlator is charge conjugation odd. It is impossible to write a function of only the momenta which is charge conjugation odd. This argument readily generalises to the vanishing of an $n$-point correlator with an odd number of spin-one current operator in the free theory.
\\[9pt]
$\mathbf{\boldsymbol \langle{T_{\mu \nu}\,J_0\,J_0} \rangle}$
\\[5pt]
The first non-trivial higher spin correlator which is of interest to us is  $\langle T_{\mu\nu}J_0\,J_0\rangle$. The stress tensor of the free bosonic theory is as given in \eqref{CCB}. As already mentioned there, we have chosen to work with a stress-tensor which is explicitly traceless.
\subsubsection*{Tensor Decomposition: Local and Transverse parts}
As discussed in \eqref{Tj0j0reconstruction} of Section \ref{reconstructionformuladetails}, the correlator can be written as the sum of a transverse part and a local part.
%Upon taking the contribution from all the relevant Wick contractions we obtain the following result which exactly matches with the expected decomposition of the full correlator into  transverse and local parts follows 
%\cite{Bzowski_2014}
%\begin{align}\label{TJ03ptb}
%\langle T_{\mu\nu}( \boldsymbol{k}_1) J_0( \boldsymbol{k}_2) J_0(( \boldsymbol{k}_3)\rangle &=\langle t_{\mu\nu}( \boldsymbol{k}_1) J_0( \boldsymbol{k}_2) J_0(( \boldsymbol{k}_3)\rangle+\langle t_{\mu\nu}^{loc}( \boldsymbol{k}_1) J_0( \boldsymbol{k}_2) J_0(( \boldsymbol{k}_3)\rangle
%\end{align}
The transverse part is given by
\begin{align}
\label{Tmunu3pt}
\langle\langle T_{\mu\nu}( \boldsymbol{k}_1) J_0( \boldsymbol{k}_2) J_0( \boldsymbol{k}_3)\rangle\rangle_{\text{transverse}}=A_{1}(k_1,k_2,k_3)\,\Pi^{\alpha \beta}_{\mu \nu}\left(\boldsymbol{k}_{1}\right) k_{2\alpha}\,k_{2\beta}\,,
\end{align}
where the transverse-traceless projector $\Pi_{\alpha \beta}^{\mu \nu}(\boldsymbol{p})$ is defined in \eqref{projk1k1}. By performing the integrals arising from Wick contractions we determined the form factor $A_{1}(k_1,k_2,k_3)$ :
\begin{align}\label{AT3pt}
A_{1}(k_1,k_2,k_3)=\frac{(2k_1+k_2+k_3)}{4 k_2 k_3(k_1+k_2+k_3)^2}\,.
\end{align}
This matches the result obtained by solving the primary conformal Ward identities \cite{Bzowski_2014} :
\begin{align}
\mathrm{K}_{ij} A_{1}=0,\quad i,j=1,2,3\,,
\end{align}
where $K_{ij}\equiv K_i-K_j$ and 
\begin{align}
\mathrm{K}_{i}=\frac{\partial^{2}}{\partial k_{i}^{2}}+\frac{d+1-2 \Delta_{i}}{k_{i}} \frac{\partial}{\partial k_{i}} \,.
%\mathrm{K}_{i j}&=\mathrm{K}_{i}-\mathrm{K}_{j}
\end{align}
Since the stress-tensor is traceless, the local part of the correlator \eqref{localtermTJ0J0}  is modified by the following additional term :
\begin{align}\label{lcTadditional}
\langle\langle &T_{\mu\nu}(\boldsymbol{k}_1) J_0( \boldsymbol{k}_2) J_0( \boldsymbol{k}_3)\rangle\rangle_{\text{additional}}=
%\frac{2}{k_1^2}\bigg[-k_{(1\mu}k_{2\nu)}\langle J_0( \boldsymbol{k}_3)J_0(- \boldsymbol{k}_3)\rangle-k_{(1\mu}k_{3\nu)}\langle J_0( \boldsymbol{k}_2)J_0(- \boldsymbol{k}_2)\rangle\bigg]\notag\\&+\pi_{\mu\nu}( \boldsymbol{k}_1)\bigg[\langle J_0( \boldsymbol{k}_2)J_0(- \boldsymbol{k}_2)\rangle+\langle J_0( \boldsymbol{k}_3)J_0(- \boldsymbol{k}_3)\rangle\bigg]\notag\\&
%-\frac{1}{2}\bigg(g_{\mu\nu}+\frac{k_{1\mu}k_{1\nu}}{k_1^2}\bigg)\bigg[-\frac{k_1.k_2}{k_1^2}\langle J_0( \boldsymbol{k}_3)J_0(- \boldsymbol{k}_3)\rangle - \frac{k_1.k_3}{k_1^2} \langle J_0( \boldsymbol{k}_2)J_0(- \boldsymbol{k}_2)\rangle\bigg]\notag\\&
-\frac{d-\Delta}{d}g_{\mu\nu}\bigg[\langle J_0( \boldsymbol{k}_2)J_0(- \boldsymbol{k}_2)\rangle+\langle J_0( \boldsymbol{k}_3)J_0(- \boldsymbol{k}_3)\rangle\bigg]\,.
\end{align}
with $d=3$ and $\Delta=1$. This ensures that the trace Ward identity \eqref{traceWI3pt} gets modified to give a zero on the R.H.S. The additional term \eqref{lcTadditional} modifies the transverse Ward identity \eqref{TWITJ0J0} by the following additional term : 
\begin{align}
k_1^\mu\langle\langle T_{\mu\nu}( \boldsymbol{k}_1) J_0(\boldsymbol{k}_2) J_0(\boldsymbol{k}_3)\rangle\rangle_{\text{additional}} &=-\frac{d-\Delta}{d}k_{1\nu}\bigg[\langle J_0( \boldsymbol{k}_2)J_0(- \boldsymbol{k}_2)\rangle+\langle J_0(\boldsymbol{k}_3)J_0(- \boldsymbol{k}_3)\rangle\bigg]\,.
%\cr
%k_1^\nu\langle T_{\mu\nu}( \boldsymbol{k}_1) J_0( \boldsymbol{k}_2) J_0(\boldsymbol{k}_3)\rangle_{\text{additional}} &=-\frac{d-\Delta}{d}k_{1\mu}\bigg[ \langle J_0( \boldsymbol{k}_2)J_0(- \boldsymbol{k}_2)\rangle+\langle J_0( \boldsymbol{k}_3)J_0(- \boldsymbol{k}_3)\rangle\bigg]\
\end{align}
and similarly when contracted with $k_1^\nu$.
%It may be easily verified that the above function $A_{1}(k_1,k_2,k_3)$ in \eqref{AT3pt} obeys the primary Ward identities derived in \cite{Bzowski_2014}
%\begin{align}
%\mathrm{K}_{ij} A_{1}=0\quad i,j=1,2,3
%\end{align}
%where $K_{ij}\equiv K_i-K_j$ and 
%\begin{align}
%\mathrm{K}_{j}=\frac{\partial^{2}}{\partial k_{j}^{2}}+\frac{d+1-2 \Delta_{j}}{k_{j}} \frac{\partial}{\partial k_{j}} 
%%\mathrm{K}_{i j}&=\mathrm{K}_{i}-\mathrm{K}_{j}
%\end{align}

Interestingly, the transverse part of the correlator in \eqref{Tmunu3pt} in three dimensions can also be expressed in terms of the projector $\chi_{\alpha \beta}^{\mu \nu}( \boldsymbol{p})$ composed of Levi-Civita tensor \eqref{Ek1k1} :
\begin{align}
\langle\langle T_{\mu\nu}( \boldsymbol{k}_1) J_0(\boldsymbol{k}_2) J_0(\boldsymbol{k}_3)\rangle\rangle_{\text{transverse}}=\chi^{\alpha \beta}_{\mu \nu}( \boldsymbol{k}_1) \widetilde{A}_{1}(k_1,k_2,k_3)\,k_{2\alpha}\,k_{2\beta}\,.
\label{tep3}
\end{align}
In this basis, the form factor $\widetilde{A}_{1}(k_1,k_2,k_3)$ is given by
\begin{align}
\widetilde{A}_{1}(k_1,k_2,k_3)=-\frac{(2k_1+k_2+k_3)}{4 k_1^2k_2 k_3(k_1+k_2+k_3)^2}\,.
\end{align}
%Let us note the following relation between the form factors $A_1$ and $\widetilde A_1$ 
%\begin{equation}
%\frac{A_{1}(k_1,k_2,k_3)}{\widetilde{A}_{1}(k_1,k_2,k_3)}=-k_1^2,
%\end{equation}
%which readily follows from \eqref{rlation1}.\\
%%
$\mathbf{\boldsymbol \langle J_\mu\,J_\nu\, J_0\rangle}$
\\[5pt]
%\subsubsection{$\boldsymbol{\langle J_\mu J_\nu J_0\rangle}$}
We now present the results for the correlator $\langle J_\mu\, J_\nu\, J_0\rangle$. 
%Details of the computation are given in Appendix \ref{Wick contractions}.
\subsubsection*{Tensor Decomposition: Local and Transverse parts}
The result for the correlator obtained by explicit integration can be expressed as the sum of a transverse part and a local part \eqref{jmujnu3pful}.
%exactly matches with the expected expression given in which states that the full correlator can be constructed from  transverse part as follows \cite{Bzowski_2014}
%
%\begin{align}\label{jmujnu3pful}
%\langle J_{\mu}( \boldsymbol{k}_1) J_\nu( \boldsymbol{k}_2) J_0(( \boldsymbol{k}_3)\rangle&=\langle j_{\mu}( \boldsymbol{k}_1) j_\nu( \boldsymbol{k}_2) J_0(( \boldsymbol{k}_3)\rangle+\langle j_{\mu}( \boldsymbol{k}_1) j_\nu( \boldsymbol{k}_2) J_0(( \boldsymbol{k}_3)\rangle^{loc}
%\end{align}
The transverse part is given by
\begin{align}\label{jmujnu3p}
\langle\langle J_{\mu}( \boldsymbol{k}_1) J_\nu( \boldsymbol{k}_2) J_0(\boldsymbol{k}_3)\rangle\rangle_{\text{transverse}}=\pi^{\alpha}_{\mu}\left(\boldsymbol{k}_{1}\right) \pi^{\beta}_{\nu}\left(\boldsymbol{k}_{2}\right)\left[A_{1}(k_1,k_2,k_3) k_{2\alpha} k_{3\beta}+A_{2}(k_1,k_2,k_3) g_{\alpha \beta}\right]
\end{align}
where the projector $\pi_{\alpha}^{\mu}\left(\boldsymbol{k}_{1}\right)$ is defined in \eqref{projk1}. By explicit computation we determined the form factors turn to be :
\begin{align}
\label{A1A2JmuJnuboson}
A_{1}(k_1,k_2,k_3)=\frac{1}{4 k_3 (k_1 + k_2 + k_3)^2},\quad
A_{2}(k_1,k_2,k_3)=\frac{1}{4  (k_1 + k_2 + k_3)}\,.
\end{align}
%\begin{align}
%\langle j_{\mu}( \boldsymbol{k}_1) j_\nu( \boldsymbol{k}_2) J_0(( \boldsymbol{k}_3)\rangle^{loc}=&\frac{k_{1\mu}k_{1\nu}}{k_1^2}\langle J_0( \boldsymbol{k}_3)J_0(- \boldsymbol{k}_3)\rangle+\frac{k_{2\mu}k_{2\nu}}{k_2^2}\langle J_0( \boldsymbol{k}_3)J_0(- \boldsymbol{k}_3)\rangle\notag\\&-\frac{k_{1\mu}k_{2\nu} k_1.k_2}{k_1^2k_2^2}\langle J_0( \boldsymbol{k}_3)J_0(- \boldsymbol{k}_3)\rangle
%\end{align}
%Notice that upon dotting with $k_{1\mu}$ and $k_{2\nu}$ the transverse part vanishes and we obtain the  Ward Identities as eq.(\ref{jmujnu3pful})
%\begin{align}
%k_1^\mu\langle J_{\mu}( \boldsymbol{k}_1) J_\nu( \boldsymbol{k}_2) J_0(( \boldsymbol{k}_3)\rangle=&k_{1\nu}\langle J_0( \boldsymbol{k}_3)J_0(- \boldsymbol{k}_3)\rangle
%\end{align}
%\begin{align}
%k_2^\nu\langle J_{\mu}( \boldsymbol{k}_1) J_\nu( \boldsymbol{k}_2) J_0(( \boldsymbol{k}_3)\rangle=&k_{2\nu}\langle J_0( \boldsymbol{k}_3)J_0(- \boldsymbol{k}_3)\rangle 
%\end{align}
This matches the result obtained in \cite{Bzowski_2014} by solving the primary Ward identities :
\begin{align}
&\mathrm{K}_{12} A_{1}=0, \quad \mathrm{K}_{13} A_{1}=0,\quad
\mathrm{K}_{12} A_{2}=0, \quad \mathrm{K}_{13} A_{2}=2 A_{1}\,.
\end{align}
The local part of the correlator is as given in \eqref{jmujnuj0local}.  This satisfies the transverse Ward identities in \eqref{transverseWIfreeboson}.

The transverse part \eqref{jmujnu3p} can also be expressed in terms of the projector $\Sigma_{\mu\nu\alpha\beta}(k_1,k_2)$ \eqref{Ek1k2} composed of Levi-Civita tensors as follows :
\begin{align}\label{jmujnu3p2}
\langle\langle J_{\mu}( \boldsymbol{k}_1) J_\nu( \boldsymbol{k}_2) J_0(\boldsymbol{k}_3)\rangle\rangle_{\text{transverse}}=\Sigma_{\mu\nu\alpha\beta}(k_1,k_2)\left[\widetilde{A}_{1}(k_1,k_2,k_3) k_{2}^{\alpha} k_{3}^{\beta}+\widetilde{A}_{2}(k_1,k_2,k_3) g^{\alpha \beta}\right]
\end{align}
where the form factors $\widetilde{A}_{1}(k_1,k_2,k_3) $ and $\widetilde{A}_{2}(k_1,k_2,k_3) $ are given by :
\begin{align}
\widetilde{A}_{1}(k_1,k_2,k_3)&=\frac{k_1^2+2 k_1 k_3+k_2^2+2 k_2 k_3+k_3^2}{8 k_1^2 k_2^2 k_3 (k_1+k_2+k_3)^2}\nonumber\\[5pt]
\widetilde{A}_{2}(k_1,k_2,k_3)&=\frac{-k_1^3+k_1^2 (k_2-k_3)+k_1 (k_2-k_3)^2-(k_2-k_3) (k_2+k_3)^2}{16 k_1^2 k_2^2 k_3 (k_1+k_2+k_3)}\,.
\end{align}
\subsubsection{Four-point functions}
We will now describe the results of our  computation of various four-point functions in the free bosonic theory.  In order to evaluate the integrals, we utilise  two interesting techniques. The first technique involves inversion of all momenta appearing in the integral (see Appendix \ref{ScInts} and Section 5.3.2 of \cite{Anninos:2017eib} and \cite{Yacoby:2018yvy}). This makes the computation of the three dimensional box integrals in momentum space easy and efficient. The second method incorporates the Schouten identity (see \eqref{lmu} of Appendix \ref{2tIntegral} and also \cite{vanNeerven:1983vr}) to simplify complicated integrals into simpler known integrals. This method turns out to be more efficient than the usual PV reduction scheme.
\subsubsection*{$\langle \boldsymbol{J_0\,J_0\,J_0\,J_0} \rangle $ }
%

%\begin{align}
%\langle& J_0(k_1) J_0(k_2) J_0(k_3) J_0(k_4)\rangle=G_{12}+G_{34}+G_{56}
%\end{align}
% Bedhotiya:2015uga
The four-point correlator of the scalar operator $J_0$ is obtained by three relevant Wick contractions followed by the use of inversion technique to evaluate the integrals. This was determined in a specific kinematic regime in \cite{Bedhotiya:2015uga,Turiaci:2018nua}. The final result is as follows :
\begin{align}
&\langle\langle J_0(\boldsymbol{k}_1) J_0(\boldsymbol{k}_2) J_0(\boldsymbol{k}_3) J_0(\boldsymbol{k}_4)\rangle\rangle\notag\\&=\frac{1}{4k_1\,k_2\,k_3\,k_{4}}\bigg[\frac{k_1\,k_{12}\,k_2+k_1\,k_{4}\,k_{23}+k_{12}\,k_{4} k_3 +k_2\,k_{23}\,k_3}{k_{12}\,k_{23} \,(k_{4}\,k_2 + k_{12}\,k_{23} + k_1\,k_3)}\bigg]+(2\leftrightarrow3 )+(3\leftrightarrow4)\,.\label{FBsc4}
\end{align}
In the above equation $(2\leftrightarrow 3)$ and $(3\leftrightarrow 4)$ denote that these terms are related to the first term by the momentum exchange $k_2\leftrightarrow k_3$ and $k_3\leftrightarrow k_4$ respectively. 
%%%%
\subsubsection*{$\langle \boldsymbol{T_{\mu\nu}\,J_0\,J_0\,J_0} \rangle$ }
\label{secJ2J0J0J0}
In the computation of $\langle T_{\mu\nu}J_0J_0J_0\rangle$  we follow \cite{corian2019fourpoint} closely. The computational details are given in Appendix \ref{WCJ24}.
The transverse part of the correlator is expressed in terms of the projector as follows :
\begin{align}
\label{pinta}
\hspace{-.5cm}\langle\langle& T^{\mu\nu}( \boldsymbol{k}_1) J_0( \boldsymbol{k}_2) J_0(\boldsymbol{k}_3)J_0( \boldsymbol{k}_4)\rangle\rangle_{\text{transverse}}=\notag\\&\Pi^{\mu \nu }_{\alpha \beta }(k_1)\left[A(k_1,k_2,k_3,k_4)\, k_2^{\alpha}\,k_3^{\beta}+B(k_1,k_2,k_3,k_4)\,k_2^{\alpha}\,k_2^{\beta}+C(k_1,k_2,k_3,k_4)\,k_3^{\alpha}\,k_3^{\beta}\right]\,.
\end{align}
where the transverse-traceless  projector $\Pi^{\mu \nu }_{\alpha \beta }(k_1)$  is given by \eqref{projk1k1}. Exchange symmetry $k_2\leftrightarrow k_3,~ k_2\leftrightarrow k_4$ and $k_3\leftrightarrow k_4$  dictates that the form factor $A(k_1,k_2,k_3,k_4)$ is the only independent one of the three in \eqref{pinta}.
One may express the other two form factors in terms of $A$ as follows :
\begin{align}\label{Symcons}
B(k_1,k_2,k_3,k_4)&=\frac{A(k_1,k_2,k_3,k_4)+A(k_1,k_4,k_3,k_2)}{2}\cr
C(k_1,k_2,k_3,k_4)&=\frac{A(k_1,k_2,k_3,k_4)+A(k_1,k_2,k_4,k_3)}{2}\,.
\end{align}
Explicit computation gives :
\begin{align}
A&=\frac{1}{4 \epsilon^{k_1k_2k_3}\epsilon_{k_1k_2k_3}} \bigg[\frac{(-\widetilde{W}^{\lambda}\epsilon_{\lambda k_1 k_3}+2 b_1(k_3^2+k_{1}\cdot k_{3})}{k_{123}\,k_{13}\,k_{3} (k_1\,k_{123} + k_{12}\,k_{13} + k_2\,k_3)}+\frac{\widetilde{W}^{\lambda}\epsilon_{\lambda k_1 k_3}+2 b_1\,k_{2}\cdot k_{3}}{k_{123}\,k_{23}k_{3} (k_{1}\,k_{3}+k_{12}\,k_{23}+k_{123}\,k_{2})}\cr
&\hspace{3cm}+\frac{(k_{123}\,k_{23}+k_{13}\,k_{3}) (\widetilde{W}^{\lambda}\epsilon_{\lambda k_1 k_3}-2\,b_{2}\,k_{2}\cdot k_{3})}{k_{123}\,k_{13}\,k_{2}\,k_{23}\,k_{3} (k_{1}\,k_{2}+k_{123}\,k_{3}+k_{13}\,k_{23})}+\frac{b_1-b_3}{k_{123}\,k_{13}\,k_{2}}+\frac{b_3}{k_{2}\,k_{23}\,k_{3}}\bigg]\,,\label{AT4pt}
\end{align}
where $\widetilde W^\mu$ and the $b_i$ are as follows :
\begin{align}\label{Wtilde}
\widetilde{W}^{\mu}&=\epsilon^{\mu k_2 k_3}k_1^2-\epsilon^{k_1\mu k_3} k_2^2-\epsilon^{k_1 k_2\mu}k_3^2\nonumber\\[5pt]
b_1=-\epsilon^{\lambda k_1 k_2}\epsilon_{\lambda k_1 k_3},\quad
&b_2=-\epsilon^{\lambda k_1 k_3}\epsilon_{\lambda k_1 k_3},\quad
b_3=-\epsilon^{\lambda k_2 k_3}\epsilon_{\lambda k_1 k_3}\,.
\end{align}
We have checked that the results for $B$ and $C$ from explicit computation satisfy \eqref{Symcons}.  

\subsubsection*{Local Part and Ward Identity}
As the stress-tensor is traceless, the local part of the correlator \eqref{localtermTJ0J0J0} gets the following additional contribution :
\begin{align}
\langle\langle& T^{\mu\nu}(\boldsymbol{k}_1) J_0( \boldsymbol{k}_2) J_0( \boldsymbol{k}_3)J_0( \boldsymbol{k}_4)\rangle\rangle_{\text{additional}}=
%\frac{2}{k_1^2}\bigg[-k_{(1\mu}k_{2\nu)}\langle J_0( \boldsymbol{k}_3+ \boldsymbol{k}_4)J_0( \boldsymbol{k}_3)J_0( \boldsymbol{k}_4)\rangle\notag\\&-k_{(1\mu}k_{3\nu)}\langle J_0( \boldsymbol{k}_2+ \boldsymbol{k}_4)J_0( \boldsymbol{k}_2)J_0( \boldsymbol{k}_4)-k_{(1\mu}k_{4\nu)}\langle J_0( \boldsymbol{k}_2+ \boldsymbol{k}_3)J_0( \boldsymbol{k}_2)J_0( \boldsymbol{k}_3)\rangle\bigg]\notag\\&+\frac{d-\Delta}{d-1}\pi^{\mu\nu}( \boldsymbol{k}_1)\bigg[\langle J_0( \boldsymbol{k}_3+ \boldsymbol{k}_4)J_0( \boldsymbol{k}_3)J_0( \boldsymbol{k}_4)\rangle+\langle J_0( \boldsymbol{k}_2+ \boldsymbol{k}_4)J_0( \boldsymbol{k}_2)J_0( \boldsymbol{k}_4)\rangle\notag\\&+\langle J_0( \boldsymbol{k}_2+ \boldsymbol{k}_3)J_0( \boldsymbol{k}_2)J_0( \boldsymbol{k}_3)\rangle\bigg]\notag\\&
%-\frac{1}{(d-1)}\bigg(g_{\mu\nu}+(d-2)\frac{k_{1\mu}k_{1\nu}}{k_1^2}\bigg)\bigg[-\frac{ \boldsymbol{k}_1. \boldsymbol{k}_2}{k_1^2}J_0( \boldsymbol{k}_3+ \boldsymbol{k}_4)J_0( \boldsymbol{k}_3)J_0( \boldsymbol{k}_4)\rangle\notag\\&- \frac{ \boldsymbol{k}_1. \boldsymbol{k}_3}{k_1^2}\langle J_0( \boldsymbol{k}_2+ \boldsymbol{k}_4)J_0( \boldsymbol{k}_2)J_0( \boldsymbol{k}_4)\rangle- \frac{ \boldsymbol{k}_1. \boldsymbol{k}_4}{k_1^2}\langle J_0( \boldsymbol{k}_2+ \boldsymbol{k}_3)J_0( \boldsymbol{k}_2)J_0( \boldsymbol{k}_3)\rangle\bigg]
-g^{\mu\nu}\bigg[\langle\langle J_0( \boldsymbol{k}_1+ \boldsymbol{k}_2)J_0( \boldsymbol{k}_3)J_0( \boldsymbol{k}_4)\rangle\rangle\cr
&\hspace{1cm}+\langle\langle J_0( \boldsymbol{k}_1+ \boldsymbol{k}_3)J_0( \boldsymbol{k}_2)J_0( \boldsymbol{k}_4)\rangle\rangle+\langle\langle J_0( \boldsymbol{k}_1+ \boldsymbol{k}_4)J_0( \boldsymbol{k}_2)J_0( \boldsymbol{k}_3)\rangle\rangle\bigg]\frac{d-\Delta}{d}\,,
\end{align}
so that the trace Ward identity \eqref{traceWI4pt} is modified to give a zero on the R.H.S.
The transverse Ward identities are modified accordingly by the following terms : 
%%%
\begin{align}
&k_{1\mu}\langle\langle T^{\mu\nu}( \boldsymbol{k}_1) J_0( \boldsymbol{k}_2) J_0( \boldsymbol{k}_3)J_0( \boldsymbol{k}_4)\rangle\rangle_{\text{additional}}\cr
&=-\frac{d-\Delta}{d}k_1^{\nu}\bigg[\langle\langle J_0( \boldsymbol{k}_1+ \boldsymbol{k}_2)J_0(\boldsymbol{k}_3)J_0( \boldsymbol{k}_4)\rangle\rangle+\langle\langle J_0( \boldsymbol{k}_1+ \boldsymbol{k}_3)J_0( \boldsymbol{k}_2)J_0( \boldsymbol{k}_4)\rangle\rangle\cr
&\hspace{1cm}+\langle\langle J_0( \boldsymbol{k}_1+ \boldsymbol{k}_4)J_0( \boldsymbol{k}_2)J_0( \boldsymbol{k}_3)\rangle\rangle\bigg]\,,
\end{align}
where the scalar three-point function is as given in \eqref{j0j0j03pt}. A similar term exists for the transverse Ward identity obtained by the contraction of the correlator with $k_{1,\nu}$.
\subsubsection*{$\langle \boldsymbol{J_\mu\,J_\nu\,J_0\,J_0} \rangle$}
\label{secJ1J1J0J0}
%We now compute the correlator $\langle J_1J_1J_0J_0\rangle $. After performing the relevant Wick contractions provided in the appendix \ref{WCJ14} we obtain
%\begin{align}
%\langle J_{\mu}( \boldsymbol{k}_1) J_\nu( \boldsymbol{k}_2) J_0(( \boldsymbol{k}_3J_0(( \boldsymbol{k}_4)\rangle=G_{12}^{\mu\nu}+G_{34}^{\mu\nu}+G_{56}^{\mu\nu}
%\end{align}
%\begin{align}
%G_{12}^{\mu\nu}&=\int\frac{2(2l+k_1)^{\mu}(2l-k_2)^{\nu}}{l^2(l+k_1)^2(l-k_2)^2(l-k_2-k_3)^2}
%\end{align}
%\begin{align}
%G_{34}^{\mu\nu}&
%=\int\frac{2(2l+k_1)^{\mu}(2l-k_2)^{\nu}}{l^2(l+k_1)^2(l-k_2)^2(l-k_2-k_4)^2}
%\end{align}
%\begin{align}
%G_{56}^{\mu\nu}&=\int\frac{2(2l+k_1)^{\mu}(2l-(2k_3+k_2))^{\nu}}{l^2(l+k_1)^2(l-k_3)^2(l-k_2-k_3)^2}
%\end{align}
%
We express the correlator as the sum of transverse and local parts \eqref{jmujnu3pful}.
The transverse part in terms of the projector takes the following form :
\begin{align}
\hspace{-1cm}\langle\langle J_{\mu}( \boldsymbol{k}_1) J_\nu( \boldsymbol{k}_2) J_0( \boldsymbol{k}_3)J_0( \boldsymbol{k}_4)\rangle\rangle_{\text{transverse}}&=\zeta^{\alpha \beta}_{\mu \nu}(\boldsymbol{k}_{1},\boldsymbol{k}_{2})\bigg[A_{1}(k_1,k_2,k_3,k_4) k_{2\alpha} k_{1\beta}+B_{1}(k_1,k_2,k_3,k_4)  k_{3\alpha} k_{1\beta}\notag\cr
&\hspace{2cm}+C_{1}(k_1,k_2,k_3,k_4) k_{2\alpha} k_{3\beta} +D_{1}(k_1,k_2,k_3,k_4) k_{3\alpha} k_{3\beta} \bigg]
\end{align}
where $\zeta_{\alpha \beta}^{\mu \nu}$ is the projector defined in \eqref{Projp1p2}.
Using exchange symmetry $k_3\leftrightarrow k_4$ and $\mu \leftrightarrow \nu, ~k_1\leftrightarrow k_2$ it is easy to see that $D_1$ and $B_1$ are determined in terms of $C_1$ as below :
\begin{align}
D_1(k_1, k_2, k_4, k_3) &= C_1(k_1, k_2, k_3, k_4) + C_1(k_1, k_2, k_4, k_3) \notag\\
B_1(k_1, k_2, k_3, k_4) &= C_1(k_2, k_1, k_3, k_4)\,.
%A_1(k_1, k_2, k_3, k_4) &=A_1(k_2, k_1, k_3, k_4) \notag\\
%D_1(k_1, k_2, k_3, k_4)&=D_1(k_1, k_2, k_4, k_3) 
\end{align}
Thus the independent form factors are $A_1$ and $C_1$. Upon explicit computation, we have :
\begin{align}
&A_1(k_1,k_2,k_3,k_4)=\frac{1}{4 \epsilon^{ k_1 k_2 k_3}\epsilon_{ k_1 k_2 k_3}}\notag\\
&\bigg[\frac{(k_{1}k_{12}+k_{123}\,k_{13}) \left(-\widetilde{W}^{\lambda}\epsilon_{\lambda k_1 k_3}+2 b_1 (k_3^2+k_{1}\cdot k_{3})\right)}{ k_{1}k_{12}\,k_{123}\,k_{13}\,k_{3}(k_{1}\,k_{123}+k_{12} k_{13}+k_{2}\,k_{3})}-\frac{\widetilde{W}^{\lambda}\epsilon_{\lambda k_1 k_3}+2 b_1 k_{2}\cdot k_{3}}{ k_{1}k_{12} k_{123}(k_{1}k_{3}+k_{12}\,k_{23}+k_{123}k_{2})}\notag\\&\hspace{3cm}-\frac{\widetilde{W}^{\lambda}\epsilon_{\lambda k_1 k_3}-2 b_{2} k_2\cdot k_3}{k_{1}k_{123}\,k_{13} (k_{1}k_{2}+k_{123}\,k_{3}+k_{13}\,k_{23})}+\frac{b_{2}+b_3}{k_{12} k_{123}\,k_{3}}-\frac{b_{2}}{ k_{1}\,k_{13}\,k_{3}}\bigg]\,,
\end{align}
where the vector $\widetilde{W}^{\mu}$ and $b_i$'s are as in \eqref{Wtilde}. Interestingly, the form factor $C_1(k_1,k_2,k_3,k_4)$ can be related to the form factor $A(k_1,k_2,k_3,k_4)$  which appeared in the previous subsection in \eqref{AT4pt} as follows :
\begin{align}
C_1(k_1,k_2,k_3,k_4)=
A(k_1,k_2,k_3,k_4)-\frac{1}{2k_{123}\,k_2\,k_{23} (k_1\,k_2 + k_{13}\,k_{23} + k_{123}\,k_3)}\,.
\end{align}
%Quite interestingly the coefficient $C_1$ is related to the corresponding form factor in the $\langle T_{\mu \nu} J_0 J_0 J_0\rangle$ correlator, $A(k_1,k_2,k_3,k_4)$ which we obtained in \eqref{AT4pt}. 
The local part of the correlator is in \eqref{j14pt}. It satisfies the transverse Ward identities \eqref{transverseWIJmuJnuJoJo}.
%Upon dotting with $k_{1\mu}$ and $k_{2\nu}$ in eq.(\ref{J14pt})  the transverse part vanishes and whereas the local part given by eq.(\ref{j14pt})  leads to the expected Ward identities 
%\begin{align}
%k_1^{\mu}\langle J_{\mu}( \boldsymbol{k}_1) J_\nu( \boldsymbol{k}_2) J_0(( \boldsymbol{k}_3J_0(( \boldsymbol{k}_4)\rangle=k_{1\nu}\langle J_0( \boldsymbol{k}_3+ \boldsymbol{k}_4)J_0( \boldsymbol{k}_3)J_0( \boldsymbol{k}_4)\rangle\notag\\
%k_2^{\nu}\langle J_{\mu}( \boldsymbol{k}_1) J_\nu( \boldsymbol{k}_2) J_0(( \boldsymbol{k}_3J_0(( \boldsymbol{k}_4)\rangle=k_{2\mu}\langle J_0( \boldsymbol{k}_3+ \boldsymbol{k}_4)J_0( \boldsymbol{k}_3)J_0( \boldsymbol{k}_4)\rangle
%\end{align}
%
\subsubsection{Five-point functions}

\subsubsection*{$\langle \boldsymbol{ J_0\,J_0\,J_0\,J_0\,J_0}\rangle$}
In this subsection we describe the result for the five-point function of the scalar operator in the free bosonic theory.  We provide the details in Appendix \ref{j0B5}. Although the method of inversion of momenta  proved to be efficient for the computation of scalar four-point correlators, it turns out to be a bit tedious for the case of five-point functions. The technique of utilising the Schouten identity turns out to be much more efficient and insightful for five-point correlators as was demonstrated in \cite{vanNeerven:1983vr}. We give the details in Appendix \ref{j0B5}.  The Wick contractions give :
\begin{align}\label{5pt}
\langle\langle &J_{0}( \boldsymbol{k}_1) J_0( \boldsymbol{k}_2) J_0(\boldsymbol{k}_3)J_0(\boldsymbol{k}_4)J_0(\boldsymbol{k}_5)\rangle\rangle\notag\\&=2 N_{B} \left(H\left(k_{1}, k_{2}, k_{3}, k_{4}, k_{5}\right)+H\left(k_{1}, k_{2}, k_{3}, k_{5}, k_{4}\right)+H\left(k_{1}, k_{2}, k_{4}, k_{3}, k_{5}\right)\right.\notag\\ &\hspace{.5cm}+H\left(k_{1}, k_{2}, k_{4}, k_{5}, k_{3}\right)+H\left(k_{1}, k_{2}, k_{5}, k_{3}, k_{4}\right)+H\left(k_{1}, k_{2}, k_{5}, k_{4}, k_{3}\right) \notag\\ &\hspace{.5cm}+H\left(k_{1}, k_{3}, k_{2}, k_{4}, k_{5}\right)+H\left(k_{1}, k_{3}, k_{2}, k_{5}, k_{4}\right)+H\left(k_{1}, k_{4}, k_{2}, k_{3}, k_{5}\right)\notag \\ &\hspace{.5cm}\left.+H\left(k_{1}, k_{4}, k_{2}, k_{5}, k_{3}\right)+H\left(k_{1}, k_{5}, k_{2}, k_{3}, k_{4}\right)+H\left(k_{1}, k_{5}, k_{2}, k_{4}, k_{3}\right)\right)\,, 
\end{align}
where
\begin{align}
H\left(k_{1}, k_{2}, k_{3}, k_{4}, k_{5}\right) \equiv \int \frac{d^{3} l}{(2 \pi)^{3}} \frac{1}{l^{2}\left(l+k_{1}\right)^{2}\left(l+k_{1}+k_{2}\right)^{2}\left(l-k_{4}-k_{5}\right)^{2}\left(l-k_{5}\right)^{2}}\,.
\end{align}
The details of the computation of the above integral are provided in Appendix \ref{j0B5}. We describe the final result here :
\begin{align}
H(k_{1}, k_{2}, k_{3}, k_{4}, k_{5})=F(k_1,k_{12},k_{123},k_{1234})\,,\label{HF1}
\end{align}
where the function $F$ is expressed in terms of integrals appearing in lower point functions which we have denoted using the letter $E$ :
\begin{align}\label{E51}
F(p_1,p_2,p_3,p_4,p_5)&=\frac{1}{f }\bigg[\epsilon^{p_1p_2p_3}(E_{0123}-E_{1234})-\epsilon^{p_2p_3p_4}(E_{0234}-E_{1234})\cr
&\hspace{1cm}+\epsilon^{p_3p_4p_1}( E_{0134}-E_{1234})-\epsilon^{p_4p_1p_2}( E_{0124}-E_{1234})\bigg]\,,
\cr
%\end{align}
%\begin{align}
&\hspace{-3cm}\text{where}\quad E_{0ijk}=\int_l \frac{1}{l^2(l+p_i)^2(l+p_j)^2(l+p_k)^2}\nonumber\\[5pt]
&\hspace{-.75cm}=\frac{1}{8 p_i\,p_j\,p_k }\frac{p_i\,p_j\,p_{ij}^{(m)}+p_j\,p_k\,p_{jk}^{(m)}+p_k\,p_i\,p_{ik}^{(m)}+p_{ij}^{(m)} p_{ik}^{(m)}\,p_{jk}^{(m)}}{p_{ij}^{(m)}p_{ik}^{(m)} p_{jk}^{(m)}(p_i\,p_{jk}^{(m)}+ p_j\,p_{ik}^{(m)}+ p_k\,p_{ij}^{(m)})}\nonumber\\[5pt]
&\hspace{-1.65cm}E_{1234}=E_{0p_{21}^{(m)}p_{31}^{(m)}p_{41}^{(m)}}\nonumber\\[5pt]
&\hspace{-1cm}f=-p_1^2\,\epsilon^{p_2p_3p_4}+p_2^2\,\epsilon^{p_3p_4p_1}-p_3^2\,\epsilon^{p_4p_1p_2}+p_4^2\,\epsilon^{p_1p_2p_3}\,.
\end{align}
Note that $E_{1234}$ is obtained by making the replacement $p_i=p_{21}^{(m)}$, $p_j=p_{31}^{(m)}$ and $p_k=p_{41}^{(m)}$ in $E_{0ijk}$. One can now make use of these expressions to determine the full five-point correlator. We emphasize that the scalar five-point function we obtained has  several interesting applications. For example, this was an important ingredient  in determining a certain beta function in  \cite{Aharony_2018} . The method we have utilised here could be generalized straightforwardly to compute more complicated correlators, such as the five-point correlator $\langle T_{\mu\nu}J_0J_0J_0J_0\rangle$.

\subsection{Free Fermionic theory in $d=3$}
We will now turn our attention to correlation functions in the free fermionic theory. As noted in Section \ref{FFTdetails}, an interesting difference from the bosonic theory is that the scalar operator $J_0$ defined in \eqref{FFS0} is parity odd. Hence the  correlation functions are structurally different from those in the bosonic theory.
Below, we present  the results for two, three and four-point functions involving various operators in the free fermionic theory.
\subsubsection{Two and Three-point functions}
\label{TwoandThreepointfunctionsFF}
The two-point function of the scalar operators is given by :
\begin{align}
\langle J_0( \boldsymbol{k})J_0(- \boldsymbol{k})\rangle=-\frac{k}{8}\,.
\end{align}
The three-point function of scalar operator $J_0(k)$ can be easily shown to be vanishing :
\begin{align}
\label{j0j0j03ptF}
\langle \langle J_0( \boldsymbol{k}_1)J_0( \boldsymbol{k}_2)J_0(\boldsymbol{k}_3)\rangle\rangle=0\,.
\end{align}
\subsubsection*{$\langle \boldsymbol{T_{\mu\nu}\,J_0\,J_0} \rangle$}
The stress tensor for the fermionic theory is as given in \eqref{FFS2}. 
The tensor decomposition of the transverse-traceless part of the correlator can be expressed as :
\begin{align}
\label{TOOTensorDecomposition}
\langle\langle T_{\mu\nu}( \boldsymbol{k}_1)\,J_0( \boldsymbol{k}_2)\,J_0( \boldsymbol{k}_3)\rangle\rangle_{\text{transverse}}&=A_1(k_1,k_2,k_3)\,\Pi_{\mu\nu}^{\alpha\beta}( \boldsymbol{k}_1)\,k_2^\alpha\,k_2^\beta\,.
\end{align}
Performing the integrals using the results in Appendix \ref{Comp}, we obtain the following :
\begin{align}
\label{formfactorTOO}
A_1(k_1,k_2,k_3)=-\frac{2k_1+k_2+k_3}{(k_1+k_2+k_3)^2}\,.
\end{align}
It can be easily checked that the form factor satisfies the dilatation Ward identity \cite{Bzowski_2014}, 
\begin{align}
\label{DWIFormFactors}
\left[2d+N_n+\sum_{j=1}^3\left(k_j\,\frac{\partial}{\partial k_j}-\Delta_j\right)\right]A_1(k_1,k_2,k_3)=0\,,
\end{align}
and the primary Ward identities \cite{Bzowski_2014},
\begin{align}
K_{ij}A_1\equiv (K_{i}-K_j)\,A_1(k_1,k_2,k_3)=0\,.
\end{align}
The local part of the correlator is given by \eqref{localtermTJ0J0} for $d=3$ and $\Delta=2$. It satisfies the transverse and trace Ward identities in \eqref{TWITJ0J0} and \eqref{traceWI3pt} respectively for $d=3$ and $\Delta=2$.
%%%
%
\subsubsection*{$\boldsymbol{\langle J_\mu\,J_\nu\,J_0\rangle}$}
The spin-one current in the free fermionic theory is as given in \eqref{FFS2}. Given the transverse Ward identities \eqref{transverseWIfermion} and the fact that the correlator is odd under parity we find it useful to write the correlator in the 
following form :
\begin{align}
\label{proposedformJmuJnuJ0}
\langle\langle J_\mu( \boldsymbol{k}_1)\,J_\nu( \boldsymbol{k}_2)\,J_0( \boldsymbol{k}_3)\rangle\rangle_{\text{transverse}}=2\Sigma^{\alpha \beta}_{\mu \nu}( \boldsymbol{k}_1, \boldsymbol{k}_2)\left[A(k_1,k_2,k_3)\,k_{1}^{\rho}+B(k_1,k_2,k_3)\,k_{2}^{\rho}\right]\epsilon_{\alpha\beta\rho}
\end{align}
where $\Sigma^{\alpha \beta}_{\mu \nu}$ is the projector we introduced in \eqref{Ek1k2}.
The presence of an odd number of Levi-Civita tensors in \eqref{proposedformJmuJnuJ0} ensures the correlator is odd under parity. Requiring the invariance of the correlator under $k_1^\mu\leftrightarrow k_2^\nu$ exchange imposes the following relation between the form factors $A$ and $B$ 
\begin{align}
\label{ABcondition}
B(k_1,k_2,k_3)=-A(k_2,k_1,k_3)\,.
\end{align}
%i
%In \eqref{proposedformJmuJnuJ0} the factor of 2 was introduced for convenience. 
An explicit computation gives : 
\begin{align}
\label{ABvalues}
A(k_1,k_2,k_3)=\frac{i}{4k_1\,(k_1+k_2+k_3)^2}\,.
\end{align}
$B(k_1,k_2,k_3)$ is obtained from $A(k_1,k_2,k_3)$ using \eqref{ABcondition}.
\subsubsection{Four-point functions}
Having obtained the three-point functions we now proceed to compute four-point functions.
\subsubsection*{$\langle \boldsymbol{J_0\,J_0\,J_0\,J_0} \rangle$}
Let us begin with simplest four-point function which is of the scalar operator $J_0$. This scalar four-point correlator was obtained in a specific kinematic regime in \cite{Bedhotiya:2015uga,Turiaci:2018nua}. The relevant Wick contractions lead to integrals of the form that are provided in Appendix \ref{Comp}. Upon utilising them we obtain the following result for the correlator :
\begin{align}\label{J04ptfff}
\langle\langle& J_{0}( \boldsymbol{k}_1) J_{0}( \boldsymbol{k}_2) J_0(\boldsymbol{k}_3)J_0(\boldsymbol{k}_4)\rangle\rangle\notag\\&=\frac{1}{4} \bigg[\frac{k_1^3-k_1\,k_{13}^2+k_1\,k_2^2-k_1\,k_{23}^2+k_1\,k_3\,k_4+k_{12}\,k_3\,k_4-k_2\,k_3\,k_4}{k_1\,k_{12}\,k_3\,k_4}\notag\\&~~~-\frac{(k_{1}-k_{23}) (2 k_{1}^2+2 k_1k_{23}-k_{12}^2-k_{13}^2+k_2^2+k_{3}^2)}{k_1k_{4}\,k_{23} (k_{1}+k_{4}+k_{23})}+\frac{\chi(k_1,k_2,k_3,k_4)}{k_1\, k_{12}\,k_{4}\,k_{23}\,k_3 (k_{4} k_2 + k_{12}\,k_{23} + k_1\,k_3)}\bigg]\notag\\&~~~+(2\leftrightarrow 3)+(3\leftrightarrow 4)\,,
\end{align}
where the function $\chi(k_1,k_2,k_3,k_4)$ is :
\begin{align}
\chi(k_1,k_2,k_3,k_4)&=k_1^2\,k_{23}\,k_3 (k_3^2-k_2^2) -k_1^3\,k_{12}(k_{23}^2-2 k_3^2)- k_1^3\,k_{4}\,k_2\,k_{23}\notag\\
&\hspace{1cm}+k_1 k_{12}\left(-k_{12}^2 k_3^2+k_{13}^2 (k_{23}^2-k_3^2)+k_2^2 (k_3^2-k_{23}^2)+k_{23}^4-2 k_{23}^2\,k_3^2+k_3^4\right)\notag\\
&\hspace{1cm}+k_1\,k_{4}\,k_2\,k_{23} (k_{13}^2-k_2^2+k_{23}^2)+ k_{12}^2\,k_{23}\,k_3(k_2^2-k_{23}^2)\cr
&\hspace{1cm}+k_2^2\,k_{23}\,k_3 (k_{13}^2-k_2^2+k_{23}^2-k_3^2)
\end{align}

%Note that $(2\leftrightarrow 3)$ and $(3\leftrightarrow 4)$ in eq.(\ref{J04ptfff}) denote that these terms are related to the first term by momentum exchange $k_2\leftrightarrow k_3$ and $k_3\leftrightarrow k_4$ respectively.

 \subsubsection*{$\langle \boldsymbol{T_{\mu\nu}\,J_0\,J_0\,J_0} \rangle$ }

Unlike the corresponding four-point correlation function in the free-bosonic theory, the free fermionic correlator $\langle {T_{\mu\nu}J_0J_0J_0} \rangle$ has only the transverse part. This is because the local part \eqref{localtermTJ0J0J0} which is composed of the three-point correlator of scalars vanishes in this theory \eqref{j0j0j03ptF}. Given that $\langle T_{\mu\nu}J_0J_0J_0\rangle$ is parity odd, the entire correlator can be expressed in terms of the parity odd projector in \eqref{Ek1k1od}  as follows :
 \begin{align}\label{J24ptfff}
 \langle\langle & T_{\mu\nu}( \boldsymbol{k}_1) J_0( \boldsymbol{k}_2) J_0(\boldsymbol{k}_3J_0(\boldsymbol{k}_4)\rangle\rangle_{\text{transverse}}\notag\\&=\Delta^{\alpha \beta}_{\mu \nu}(\boldsymbol{k}_1)\bigg[A_{2}(k_1,k_2,k_3,k_4) k_{2\alpha} k_{2\beta}+B_{2}(k_1,k_2,k_3,k_4)  (k_{2\alpha} k_{3\beta}+k_{3\alpha} k_{2\beta})+C_{2}(k_1,k_2,k_3,k_4) k_{3\alpha} k_{3\beta} \bigg]
 \end{align}
This is by construction transverse to momentum $k_1$.
The form factors are determined by performing the integrals appearing in the Wick contractions. The required integrals are once again of the form given in Appendix \ref{Comp} and we obtain :
 \begin{align}
&B_{2}(k_1,k_2,k_3,k_4)=\frac{k_{2}}{2 k_{12}\,k_{123} (k_{1}k_{123}+k_{12}\,k_{13}+k_{2} k_{3})}-\frac{k_{2} (k_{12}\,k_{2}+k_{123}\,k_{23})}{2\,k_{12}\,k_{123}\,k_{23}\,k_{3} (k_{1}k_{3}+k_{12}\,k_{23}+k_{123}\,k_{2})}\notag\\&+\frac{k_{2}^2+2\,k_{2}\cdot k_{3}}{2 k_{123}\,k_{2}\,k_{23}(k_{1}\,k_{2}+k_{123}\,k_{3}+k_{13}\,k_{23})}+\frac{1}{k_{123}\,k_{23}(k_{1}+k_{123}+k_{23})}\notag\\&-\frac{2\,k_{1}+k_{123}+2 k_{23}}{ k_{123}\,k_{23}(k_{1}+k_{123}+k_{23})^2}+\frac{k_{12}+k_{123}-k_{3}}{2 k_{12}\,k_{123}\,k_{3} (k_{12}+k_{123}+k_{3})}\notag\\&\frac{1}{\epsilon^{k_1k_2k_3}\epsilon_{k_1k_2k_3}}\Bigg[\frac{(b_1+b_2) \left(k_1\cdot k_2+k_{2}^2+k_{2}\cdot k_{3}\right)}{k_{1} k_{123}(k_{1}+k_{123}+k_{23})}-\frac{(b_1-b_3) k_{2}\cdot k_{3}}{2 k_{123}\,k_{13}\,k_{2}}-\frac{b_3 k_2\cdot k_3}{2 k_{2} k_{23}k_{3}}\notag\\&-\frac{b_1(k_1+k_3+k_{13})+b_3(k_{123}-k_3-k_{13}+k_{23})}{(k_{1}+k_{123}+k_{23}) (k_{1}+k_{13}+k_{3})}-\frac{b_1 \left(k_1\cdot k_2+k_{2}^2\right)}{k_{1}\,k_{12} (k_{1}+k_{12}+k_{2})}\notag\\&+\frac{(b_2+b_3) \left(k_{1}\cdot k_{2} (k_{12}+k_{123}-k_{3})+k_{2}^2 (k_{12}+k_{123}-k_{3})+2 k_{12}\,k_2\cdot k_3\right)}{2 k_{12}\,k_{123}k_{3} (k_{12}+k_{123}+k_{3})}\notag\\&-\frac{b_2 \left(k_{2}^2 (k_{1}+k_{13}+k_{3})+k_{1}\cdot k_{2} (k_{1}+k_{13}+k_{3})+2 (k_{1}+k_{3}) k_{2}\cdot k_{3}\right)}{2 k_{1}\,k_{13}\,k_{3} (k_{1}+k_{13}+k_{3})}\notag\\&+\frac{\left(-\widetilde{W}^{\mu}\epsilon_{\mu k_1 k_3}+2 b_1 (k_3^2+k_1\cdot k_3)\right) \left(k_{2}^2 (k_{1}\,k_{12}+k_{123}\,k_{13})+k_1\cdot k_2 (k_{1}\,k_{12}+k_{123}\,k_{13})+k_{1}\,k_{12}\,k_2\cdot k_3\right)}{2 k_{1}\,k_{12}\,k_{123}\,k_{13}\,k_{3} (k_{1} k_{123}+k_{12} k_{13}+k_{2} k_{3})}\notag\\&-\frac{\left(\widetilde{W}^{\lambda}\epsilon_{\lambda k_1 k_3}+2 b_1\,k_{2}\cdot k_{3}\right)\left(k_{1}\,k_{12}\,k_2\cdot k_3-k_{23}\,k_{3}\,k_1\cdot k_2-k_{2}^2\,k_{23}\,k_{3}\right) }{2 k_{1} k_{12} k_{123}k_{23}k_{3} (k_{1} k_{3}+k_{12} k_{23}+k_{123}k_{2})}\notag\\&-\frac{\left(\widetilde{W}^{\lambda}\epsilon_{\lambda k_1 k_3}-2 b_2\, k_2\cdot k_3\right) \left(k_2\cdot k_3 (k_{1}\,k_{123}\,k_{23}+k_{1}\,k_{13} k_{3}+2 k_{2}\,k_{23}\,k_{3})+k_{2}\,k_{23}\,k_{3}\, k_{1}\cdot k_{2}+k_{2}^3\,k_{23}\,k_{3}\right)}{2\,k_{1}\,k_{123}\,k_{13}\,k_{2}\,k_{23}\,k_{3} (k_{1} k_{2}+k_{123}\,k_{3}+k_{13}\,k_{23})}\Bigg]
 \end{align}
where the vector $\widetilde{W}^{\mu}$ and $b_i$ $(i=1,2,3)$ are those which appeared in the corresponding correlator in the bosonic theory and are given by \eqref{Wtilde}. The form factors $A_2$ and $C_2$ are determined in terms of $B_2$ as follows :
\begin{align}
A_2(k_1,k_2k_3,k_4)&=B_2(k_1,k_3,k_2,k_4)+B_2(k_1,k_3,k_4,k_2)\notag\\
C_2(k_1,k_2,k_3,k_4)&=B_2(k_1,k_2,k_3,k_4)+B_2(k_1,k_2,k_4,k_3)\,.
\end{align}
This completes our computation of the correlator $\langle{T_{\mu\nu}J_0J_0J_0} \rangle$.
%\subsubsection{Five-point function}
%\subsubsection*{$\langle \boldsymbol{J_0 J_0J_0J_0J_0^F} \rangle$}
% \textcolor{red}{The explicit computation of the trace identities of the gamma matrices could be used to demonstrate that the five-point function of $J_0$ operators in the free fermionic theory vanishes. We also emphasize that one may generalize our techniques straightforwardly to compute correlators of the sort $\langle \boldsymbol{T_{\mu\nu} J_0J_0J_0J_0^F} \rangle$.}
%%%%
%%%%%
\section{Correlators from higher spin Ward identities}
\label{CorrelatorsfromhigherspinWardidentities}
In this section we discuss how correlators in theories with a higher spin symmetry can be obtained by solving the corresponding Ward identities \cite{Giombi_2012,Aharony_2012,maldacena2011constraining,Maldacena_2013,Giombi_2017,Li:2019twz}. If we denote the charge associated to the higher spin symmetry by $Q_s$ \footnote{The charge $Q_s$ is obtained from the current $J_s$ by integrating it over a co-dimension 1 hypersurface : $Q_s=\int_{x^{+}=\text{const}}\,dx^{-}\,dy\,J_{\underbrace{--\ldots -}_{s\,\text{times}}}$. We have used the lightcone coordinates in which $ds^2=dx^+dx^-+dy^2.$}, the Ward identity takes the form : 
\begin{align}
\label{HSWI}
Q_s\langle\langle \mathcal O_1(x_1)\ldots \mathcal O_n(x_n)\rangle\rangle&=\sum_{i=1}^n\langle\langle \mathcal O_1(x_1)\ldots [Q_s,\mathcal O_i(x_i)]\ldots \mathcal O_n(x_n)\rangle\rangle\cr
&=0
\end{align}
%
%Below we consider the  action of spin-3 and spin-4 charges. 
Higher spin equations in momentum space have the nice feature that they are algebraic as opposed to their position space counterparts that are differential equations. Hence, one would expect them to be easily solvable. However, the fact that we do not know the analogue of position space conformal cross-ratios $u$ and $v$ in momentum space leads to a proliferation of unknown variables to solve for. These unknowns very often exceed the number of algebraic constraints coming from the higher spin equations. In such cases one needs to resort to conformal invariance to get additional constraints. Nevertheless, we shall see that 
we are able to solve for some three-point functions without invoking conformal invariance. At the level of four-point functions we show that higher-spin equations can be used to verify the explicit results for correlators obtained by direct computation. 

\subsection{Three-point spinning correlators from higher spin equations}
In this section we discuss how higher spin Ward identities may be put to use to compute some three-point spinning correlators in the free bosonic theory without invoking conformal invariance. In our analysis we stick to spin 2 and spin 3 charges denoted by $Q_3$ and $Q_4$ respectively. The action of these charges on the spin 0 and spin 1 operators is given by :
\begin{align}
\label{Q4Q3action}
[Q_3,J_0]&=\partial_{-}J_{-}\cr
[Q_3,J_{-}]&=4\,\partial_{-}T_{--}-\frac 12\partial_{-}^3J_0\cr
[Q_4,J_0]&=\partial_{-}^3J_0-\frac{24}{5}\partial_{-}T_{--}
\end{align}
One can check that the algebra is consistent with the data in Table.~\ref{tab1}. Especially one can see that although a dimension and spin analysis would allow for a stress-tensor term in the $[Q_3,J_0]$ algebra, charge conjugation forbids this. %Especially, $Q_3$ being odd under charge conjugation, the commutator of $Q_3$ with $J_0$  should be such that 
\begin{table}[h]
\begin{tabular}
{ |p{2cm}|p{3cm}|p{1.5cm}|p{1.7cm}|p{3.5cm}|p{2cm}|}
 \hline
% \multicolumn{4}{|c|}{Country List} \\
 %\hline
Operator & Dimension ($\Delta$) & Spin ($s$) & Twist ($\tau$) & Charge conjugation & Parity\\
 \hline
$ Q_3$   & 2    &2&   0 & odd & even\\ \hline
 $Q_4$&   3  & 3   &0 & even & even\\ \hline
 $J_0$ &1 & 0&  1 & even & even\\ \hline
 $J_1$   &2 & 1&  1 & odd &even \\ \hline
 $T_{\mu\nu}$&   3  & 2&1 & even & even\\
 \hline
 \end{tabular}
 \caption{Data of charges and operators of interest in the bosonic theory}
 \label{tab1}
 \end{table}
%
%In our computation we will also require the three-point function of scalars :
%%
%\begin{align}
%\label{J0J0J0boson}
%\langle J_{0}( \boldsymbol{k}_1)\,J_0( \boldsymbol{k}_2)\,J_0( \boldsymbol{k}_3)\rangle=\frac{1}{4k_1k_2k_3}
%\end{align}
%%
\subsubsection*{Computing $\langle T_{\mu\nu}\,J_0\,J_0\rangle$ from higher spin equations}
We will first show how to obtain the $\langle T\,J_0\,J_0\rangle$ correlator by solving a higher spin equation. The first step in this regard is to identify the higher spin charge and the correlator on which the chosen charge must act.
From the action of the charges in \eqref{Q4Q3action} it is clear that the 
Ward identity \eqref{HSWI} for the charge $Q_4$ on the correlator $\langle J_0\,J_0\,J_0\rangle$ generates $\langle T\,J_0\,J_0\rangle$. The higher spin Ward identity takes the form :
\begin{align}
\label{eqnstep1}
 0&=\langle\langle[Q_4,J_0( \boldsymbol{x}_1)]J_0( \boldsymbol{x}_2)J_0( \boldsymbol{x}_3)\rangle\rangle+\langle\langle J_0( \boldsymbol{x}_1)[Q_4,J_0( \boldsymbol{x}_2)]J_0( \boldsymbol{x}_3)\rangle\rangle+\langle\langle J_0 (\boldsymbol{x}_1) J_0( \boldsymbol{x}_2)[Q_4,J_0( \boldsymbol{x}_3)]\rangle\rangle\cr
 &=\partial_{1,-}^3\langle\langle J_0( \boldsymbol{x}_1) J_0( \boldsymbol{x}_2) J_0( \boldsymbol{x}_3)\rangle\rangle-\frac{24}{5}\partial_{1,-}\langle\langle T_{--}( \boldsymbol{x}_1)J_0( \boldsymbol{x}_2)J_0( \boldsymbol{x}_3)\rangle\rangle+\text{permutations}\,,
\end{align}
where in the second line we used the algebra in \eqref{Q4Q3action}. We now express this equation in momentum space by performing a Fourier transform :
%In momentum space, the Ward identity takes the form :
 %
 \begin{align}
 \label{WITOO}
 0&=\left(k_{1-}^3+k_{2-}^3+k_{3-}^3\right)\langle\langle J_0 ( \boldsymbol{k}_1)J_0( \boldsymbol{k}_2) J_0( \boldsymbol{k}_3)\rangle\rangle\cr
 &\hspace{.5cm}+\frac{24}{5}\big(k_{1-}\langle\langle T_{--} ( \boldsymbol{k}_1)J_0( \boldsymbol{k}_2) J_0( \boldsymbol{k}_3)\rangle\rangle+k_{2-}\langle\langle J_0( \boldsymbol{k}_1)T_{--} ( \boldsymbol{k}_2) J_0( \boldsymbol{k}_3)\rangle\rangle\cr
 &\hspace{2cm}+k_{3-}\langle\langle J_0 ( \boldsymbol{k}_1)J_0( \boldsymbol{k}_2) T_{--}( \boldsymbol{k}_3)\rangle\rangle\big)\nonumber\\[5pt]
&= \left(k_{1-}^3+k_{2-}^3+k_{3-}^3\right)\langle\langle J_0 ( \boldsymbol{k}_1)J_0( \boldsymbol{k}_2) J_0( \boldsymbol{k}_3)\rangle\rangle\cr
 &\hspace{.3cm}+\frac{24}{5}\bigg[k_{1-}(\langle\langle T_{--} ( \boldsymbol{k}_1)J_0( \boldsymbol{k}_2) J_0( \boldsymbol{k}_3)\rangle\rangle_{\text{transverse}}+\langle\langle T_{--} ( \boldsymbol{k}_1)J_0( \boldsymbol{k}_2) J_0( \boldsymbol{k}_3)\rangle\rangle_{\text{local}})\cr
 & \hspace{1.2cm}+k_{2-}(\langle\langle J_0( \boldsymbol{k}_1)T_{--} ( \boldsymbol{k}_2) J_0( \boldsymbol{k}_3)\rangle\rangle_{\text{transverse}}+\langle\langle J_0( \boldsymbol{k}_1)T_{--} ( \boldsymbol{k}_2) J_0( \boldsymbol{k}_3)\rangle\rangle_{\text{local}})\cr
 &\hspace{1.2cm}+k_{3-}(\langle\langle J_0 ( \boldsymbol{k}_1)J_0( \boldsymbol{k}_2) T_{--}( \boldsymbol{k}_3)\rangle_{\text{transverse}}+\langle\langle J_0 ( \boldsymbol{k}_1)J_0( \boldsymbol{k}_2) T_{--}( \boldsymbol{k}_3)\rangle\rangle_{\text{local}})\bigg]
 \end{align}
where the local part of the  $\langle T_{\mu\nu}\,J_0\,J_0\rangle$ correlator is given by the sum of the expressions in \eqref{localtermTJ0J0} and \eqref{lcTadditional}. Our goal is to solve \eqref{WITOO} for the form factor in the transverse-traceless part :
\begin{align}
\langle\langle T^{\mu\nu}( \boldsymbol{k}_1)\,J_0( \boldsymbol{k}_2)\,J_0( \boldsymbol{k}_3)\rangle\rangle_{\text{transverse}}&=A(k_1,k_2,k_3)\,\Pi^{\mu\nu}_{\alpha\beta}( \boldsymbol{k}_1)\,k_2^\alpha\,k_2^\beta\,.
\end{align}
Accounting for the three form factors that appear on the R.H.S of \eqref{WITOO} (one from each of the three correlators), solving \eqref{WITOO} amounts to solving three linearly independent equations in three variables. Solving them we obtain the following :
\begin{align}
\label{T00FormFactorFromHSE}
A(k_1,k_2,k_3)=\frac{2k_1+k_2+k_3}{4k_2 k_3(k_1+k_2+k_3)^2}\,.
\end{align} 
The form factors in  $\langle J_0\,T_{\mu\nu}\,J_0\rangle$ and  $\langle J_0\,J_0\,T_{\mu\nu}\rangle$ are also solved for and they are related to $A(k_1,k_2,k_3)$ in \eqref{T00FormFactorFromHSE} by $k_1\leftrightarrow k_2$ and $k_1\leftrightarrow k_3$ exchange, respectively. This matches the result obtained by direct computation in \eqref{AT3pt}.  
\subsubsection*{Computing $\langle J_\mu\,J_\nu\,J_0\rangle$ from higher spin equations}
Now that we have solved for the form factor in the $\langle T_{\mu\nu}\,J_0\,J_0\rangle$ correlator, we can use it to solve a higher spin Ward identity to obtain the form factors in the $\langle J_\mu\,J_\nu\,J_0\rangle$ correlator :
\begin{align}
\label{jmujnuj0freeboson3d}
\langle\langle J_\mu( \boldsymbol{k}_1)\,J_\nu( \boldsymbol{k}_2)\,J_0( \boldsymbol{k}_3)\rangle\rangle_{\text{transverse}}=\pi_\mu^\alpha( \boldsymbol{k}_1)\,\pi_\nu^\beta( \boldsymbol{k}_2)\,\left[A_1(k_1,k_2,k_3)\,k_{2\alpha}\,k_{3\beta}+A_2(k_1,k_2,k_3)\,g_{\alpha\beta}\right]\cr
\langle\langle J_\mu( \boldsymbol{k}_1)\,J_0( \boldsymbol{k}_2)\,J_\nu( \boldsymbol{k}_3)\,\rangle\rangle_{\text{transverse}}=\pi_\mu^\alpha( \boldsymbol{k}_1)\,\pi_\nu^\beta( \boldsymbol{k}_3)\,\left[A_3(k_1,k_2,k_3)\,k_{2\alpha}\,k_{3\beta}+A_4(k_1,k_2,k_3)\,g_{\alpha\beta}\right]
\end{align}
From \eqref{Q4Q3action} it is clear that the Ward identity \eqref{HSWI} for the charge $Q_3$ on the correlator $\langle J_{-} ( \boldsymbol{k}_1)J_0( \boldsymbol{k}_2) J_0( \boldsymbol{k}_3)\rangle$ generates the required correlator. 
%
%We start with the correlator $\langle J_{-} ( \boldsymbol{k}_1)J_0( \boldsymbol{k}_2) J_0( \boldsymbol{k}_3)\rangle$ and compute the action of $Q_3$ on it. 
Following \eqref{eqnstep1} and \eqref{WITOO} the higher spin equation takes the following form in momentum space :
\begin{align}
\label{HSWIQ3explicit}
0&=4k_{1-}\langle\langle T_{--}( \boldsymbol{k}_1)\,J_0( \boldsymbol{k}_2)\,J_0( \boldsymbol{k}_3)\rangle\rangle+\frac{k_{1-}^3}{2}\langle\langle J_{0}( \boldsymbol{k}_1)\,J_0( \boldsymbol{k}_2)\,J_0( \boldsymbol{k}_3)\rangle\rangle\cr
&\hspace{.5cm}+k_{2-}\langle\langle J_{-}( \boldsymbol{k}_1)\,J_{-}( \boldsymbol{k}_2)\,J_0( \boldsymbol{k}_3)\rangle\rangle+k_{3-}\langle\langle J_{-}( \boldsymbol{k}_1)\,J_0( \boldsymbol{k}_2)\,J_{-}( \boldsymbol{k}_3)\rangle\rangle\,.
\end{align}
%jmujnuj0local
%It is straightforWard to verify that the results obtained for $\langle T_{\mu\nu}( \boldsymbol{k}_1)\,J_0( \boldsymbol{k}_2)\,J_0( \boldsymbol{k}_3)\rangle$ and $\langle J_\mu( \boldsymbol{k}_1)\,J_\nu( \boldsymbol{k}_2)\,J_0( \boldsymbol{k}_3)\rangle$ in \eqref{ref5} and \eqref{ref6} respectively combined with the scalar three-point function \eqref{J0J0J0boson} solve the higher spin equation \eqref{HSWIQ3explicit}.
Having solved for the form factor in the transverse-traceless part of the $\langle T_{\mu\nu}\,J_0\,J_0\rangle$ correlator \eqref{T00FormFactorFromHSE}, one can combine it with the local term in $\langle T_{\mu\nu}\,J_0\,J_0\rangle$ (see \eqref{localtermTJ0J0}, \eqref{lcTadditional}), the local term in $\langle J_{\mu}\,J_{\nu}\,J_0\rangle$  \eqref{jmujnuj0local} and the scalar three-point function \eqref{j0j0j03pt} to solve four linearly independent equations from \eqref{HSWIQ3explicit} to obtain the form factors $A_1,A_2,A_3$ and $A_4$ in \eqref{jmujnuj0freeboson3d} :
%Note that the $\langle J_{\mu}\,J_{\nu}\,J_0\rangle$ correlator has no local term.
\begin{align}
A_1(k_1,k_2,k_3)&=\frac{1}{4k_3(k_1+k_2+k_3)^2},\quad A_2(k_1,k_2,k_3)=\frac{1}{4(k_1+k_2+k_3)}\,.
%A_3(k_1,k_2,k_3)&=A_1(k_1,k_3,k_2),\quad\quad A_4(k_1,k_2,k_3)=A_2(k_1,k_3,k_2)
\end{align}
The form factors $A_3(k_1,k_2,k_3)$ and $A_4(k_1,k_2,k_3)$ are also solved for and are given by a $k_2\leftrightarrow k_3$ exchange in $A_1(k_1,k_2,k_3)$ and $A_2(k_1,k_2,k_3)$ respectively. This precisely matches the results obtained in \eqref{A1A2JmuJnuboson}.

Thus we have illustrated how higher spin equations determine three-point spinning correlators in the bosonic theory without invoking conformal invariance.

\subsection{Four-point spinning correlators and higher spin equations}
In this section we extend the  analysis of the previous section to four-point functions. Let us first look at the Ward identity corresponding to the charge $Q_4$ on the $\langle J_0\,J_0\,J_0\,J_0\rangle$ correlator. Following \eqref{eqnstep1} and \eqref{WITOO} it takes the following form in momentum space :
\begin{align}
\label{L.H.SQ4boson3d}
0&=\left(k_{1-}^3+k_{2-}^3+k_{3-}^3+k_{4-}^3\right)\langle\langle J_0 ( \boldsymbol{k}_1)J_0( \boldsymbol{k}_2) J_0( \boldsymbol{k}_3)J_0( \boldsymbol{k}_4)\rangle\rangle\cr
 &\hspace{.2cm}+\frac{24}{5}\Bigg(k_{1-}\langle\langle T_{--} ( \boldsymbol{k}_1)J_0( \boldsymbol{k}_2) J_0( \boldsymbol{k}_3)J_0( \boldsymbol{k}_4)\rangle\rangle+k_{2-}\langle\langle J_0( \boldsymbol{k}_1)T_{--} ( \boldsymbol{k}_2) J_0( \boldsymbol{k}_3)J_0( \boldsymbol{k}_4)\rangle\rangle\cr
&\hspace{.4cm}+k_{3-}\langle\langle J_0 ( \boldsymbol{k}_1)J_0( \boldsymbol{k}_2) T_{--}( \boldsymbol{k}_3)J_0( \boldsymbol{k}_4)\rangle\rangle+k_{4-}\langle\langle J_0 ( \boldsymbol{k}_1)J_0( \boldsymbol{k}_2) J_0( \boldsymbol{k}_3)T_{--}( \boldsymbol{k}_4)\rangle\rangle\Bigg)
\end{align}
The transverse-traceless part of the correlator is given by
\begin{align}\label{pinta2}
\langle\langle& T^{\mu\nu}( \boldsymbol{k}_1) J_0( \boldsymbol{k}_2) J_0( \boldsymbol{k}_3)J_0(\boldsymbol{k}_4)\rangle\rangle_{\text{transverse}}=\notag\\&\Pi^{\mu \nu }_{\alpha \beta }(k_1)\left[A(k_1,k_2,k_3,k_4)\,k_2^{\alpha}\,k_3^{\beta}+B(k_1,k_2,k_3,k_4)\,k_2^{\alpha}\,k_2^{\beta}+C(k_1,k_2k_3,k_4)\,k_3^{\alpha}\,k_3^{\beta}\right]\,.
\end{align}
One can easily check that the scalar four-point function \eqref{FBsc4} and the spinning correlator $\langle T_{\mu\nu}\,J_0\,J_0\,J_0\rangle$ computed in Section \ref{secJ2J0J0J0} solve \eqref{L.H.SQ4boson3d}. However, it is not possible to solve for the form factors in $\langle T_{\mu\nu}\,J_0\,J_0\,J_0\rangle$ as it amounts to solving 10 linear equations in 12 variables. 
%One would require additional constraints coming from conformal invariance to solve for these variables.

We will now show how the form factors in the $\langle T_{\mu\nu}\, J_{0}\,J_0\,J_0\rangle$ correlator can be solved for with the additional knowledge of the spinning correlator $\langle J_{\mu}\,J_\nu\,J_0\,J_0\rangle$. For this we consider  the Ward identity corresponding to the action of $Q_3$ on $\langle J_{-}\, J_{0}\,J_0\,J_0\rangle$ :
\begin{align}
\label{HSWIQ3explicit2}
0&=4k_{1-}\langle\langle T_{--}( \boldsymbol{k}_1)\,J_0( \boldsymbol{k}_2)\,J_0( \boldsymbol{k}_3)\,J_0( \boldsymbol{k}_4)\rangle\rangle+\frac{k_{1-}^3}{2}\langle\langle J_{0}( \boldsymbol{k}_1)\,J_0( \boldsymbol{k}_2)\,J_0( \boldsymbol{k}_3)\,J_0( \boldsymbol{k}_4)\rangle\rangle\cr
&\hspace{.5cm}+k_{2-}\langle\langle J_{-}( \boldsymbol{k}_1)\,J_{-}( \boldsymbol{k}_2)\,J_0( \boldsymbol{k}_3)\,J_0( \boldsymbol{k}_4)\rangle\rangle+k_{3-}\langle\langle J_{-}( \boldsymbol{k}_1)\,J_0( \boldsymbol{k}_2)\,J_{-}( \boldsymbol{k}_3)\,J_0( \boldsymbol{k}_4)\rangle\rangle\cr
&\hspace{.5cm}+k_{4-}\langle\langle J_{-}( \boldsymbol{k}_1)\,J_0( \boldsymbol{k}_2)\,J_{0}( \boldsymbol{k}_3)\,J_{-}( \boldsymbol{k}_4)\rangle\rangle\,.
\end{align}
One can easily verify that the explicit results obtained for the scalar four-point correlator in \eqref{FBsc4}, the $\langle T_{\mu\nu}\,J_0\,J_0\,J_0\rangle$ correlator in Section \ref{secJ2J0J0J0} and the $\langle J_\mu\,J_\nu\,J_0\,J_0\rangle$ correlator in Section \ref{secJ1J1J0J0} solve \eqref{HSWIQ3explicit2}. But one can do better by actually solving for the form factors. Given the spinning correlator $\langle J_{\mu}\,J_\nu\,J_0\,J_0\rangle$ and the four-point function of scalars, the higher spin Ward identity \eqref{HSWIQ3explicit2} completely determines the form factors $A$, $B$, and $C$ that appear in the transverse-traceless part of the $\langle T_{\mu\nu}\,J_0\,J_0\,J_0\rangle$ correlator \eqref{pinta2}. However, solving for the form factors in $\langle J_{\mu}\,J_\nu\,J_0\,J_0\rangle$ given the  $\langle T_{\mu\nu}\,J_0\,J_0\,J_0\rangle$ correlator is not possible as it amounts to solving for 12 variables in 10 linear equations. 

In cases where the number of unknowns to solve for exceed the number of higher spin equations,  one requires additional input and this comes from the constraints imposed by  conformal Ward identities. In the following sub-section we illustrate this in the context of the free fermionic theory where one can see this already at the level of three-point functions.
\subsection{Free Fermionic theory}
%In the previous section we saw that the four-point spinning correlator with two spin-one currents cannot be solved for using only the higher spin Ward identity. In order to solve for the correlator one needs to use the constraints that come from imposing conformal invariance. In this section we see this in some detail in the free fermionic theory. 
In this sub-section we will  discuss how the three-point correlator $\langle T_{\mu\nu}\,J_0\,J_0\rangle$ in the free fermionic theory may be obtained by solving the higher spin Ward identity. For this we impose the Ward identity \eqref{HSWI} corresponding to the generator $Q_4$ on the correlator $\langle J_0 ( \boldsymbol{k}_1)J_0( \boldsymbol{k}_2) J_0( \boldsymbol{k}_3)\rangle$. The action of $Q_4$ on $J_0$ is given by :
\begin{align}
[Q_4,J_0]=\partial_{-}^3 J_0+\epsilon_{-\mu\nu}\partial_{-}\partial^\mu\,J^{\nu}_{-}
\end{align}
\begin{table}[h]
\begin{tabular}{ |p{2cm}|p{3cm}|p{1.5cm}|p{1.7cm}|p{3.5cm}|p{2cm}|}
 \hline
% \multicolumn{4}{|c|}{Country List} \\
 %\hline
Operator & Dimension ($\Delta$) & Spin ($s$) & Twist ($\tau$) & Charge conjugation & Parity\\
 \hline
%$ Q_3$   & 3    &2&   1 & odd & even\\ \hline
 $Q_4$&   3  & 3   &0 & even & even\\ \hline
 $J_0$ &2 & 0&  2 & even & odd\\ \hline
 $J_1$   &2 & 1&  1 & odd &even \\ \hline
 $T_{\mu\nu}$&   3  & 2&1 & even & even\\
 \hline
 \end{tabular}
  \caption{Data of charges and operators of interest in the fermionic theory}
 \label{tab2}
 \end{table}
Combined with \eqref{HSWI} we get the following higher spin equation in momentum space :
\begin{align}
\label{HSE}
-ik_{1-}^3\langle\langle J_0( \boldsymbol{k}_1)J_0( \boldsymbol{k}_2)J_0( \boldsymbol{k}_3)\rangle\rangle-\epsilon_{-\mu\nu}k_{1-}k_1^\mu\langle\langle T^\nu_-( \boldsymbol{k}_1)J_0( \boldsymbol{k}_2)J_0( \boldsymbol{k}_3)\rangle\rangle+(1\leftrightarrow 2,3)=0\,.
\end{align}
Since the three-point function $\langle J_0(k_1)J_0(k_2)J_0(k_3)\rangle$ vanishes in the free fermionic theory, the equation becomes :
\begin{align}
\label{HSEQ4FF}
&\epsilon_{-\mu\nu}\left(k_{1-}k_1^\mu\langle\langle T^\nu_-( \boldsymbol{k}_1)J_0( \boldsymbol{k}_2)J_0( \boldsymbol{k}_3)\rangle\rangle+k_{2-}k_2^\mu\langle\langle J_0( \boldsymbol{k}_1)T^\nu_-( \boldsymbol{k}_2)J_0( \boldsymbol{k}_3)\rangle\rangle\right.\nonumber\\[5pt]
&\hspace{4cm}\left.+k_{3-}k_3^\mu\langle\langle J_0( \boldsymbol{k}_1)J_0( \boldsymbol{k}_2)T^\nu_-( \boldsymbol{k}_3)\rangle\rangle\right)=0
\end{align}
One can easily check that the expression for $\langle T_{\mu\nu}\,J_0\,J_0\rangle$ (see \eqref{localtermTJ0J0} and Section \ref{TwoandThreepointfunctionsFF}) solves \eqref{HSEQ4FF}.
%Note that this requires not just the transverse-traceless part \eqref{TOOTensorDecomposition} of the correlator
%\eqref{TOOttplusloc} but also the local part \eqref{TOOLoc}. 

We now ask if we can do better and actually solve \eqref{HSEQ4FF} for the form factor in the transverse-traceless part of the  $\langle T_{\mu\nu}\,J_0\,J_0\rangle$ correlator  \eqref{TOOTensorDecomposition}.
%Let us denote the form factor in $\langle T_{\mu\nu} ( \boldsymbol{k}_1)J_0( \boldsymbol{k}_2) J_0( \boldsymbol{k}_3)\rangle$, $\langle T_{\mu\nu} ( \boldsymbol{k}_1)J_0( \boldsymbol{k}_2) J_0( \boldsymbol{k}_3)\rangle$ and $\langle T_{\mu\nu} ( \boldsymbol{k}_1)J_0( \boldsymbol{k}_2) J_0( \boldsymbol{k}_3)\rangle$ by $A_1, A_2$ and  $A_3$ respectively, i.e.
%\begin{align}
%\langle t_{\mu\nu}( \boldsymbol{k}_1)\,J_0( \boldsymbol{k}_2)\,J_0( \boldsymbol{k}_3)\rangle&=A_1(k_1,k_2,k_3)\,\Pi_{\mu\nu}^{\alpha\beta}( \boldsymbol{k}_1)\,k_2^\alpha\,k_2^\beta\cr
%\langle t_{\mu\nu}( \boldsymbol{k}_2)\,J_0( \boldsymbol{k}_1)\,J_0( \boldsymbol{k}_3)\rangle&=A_2(k_1,k_2,k_3)\,\Pi_{\mu\nu}^{\alpha\beta}( \boldsymbol{k}_2)\,k_1^\alpha\,k_1^\beta\cr
%\langle t_{\mu\nu}( \boldsymbol{k}_3)\,J_0( \boldsymbol{k}_2)\,J_0( \boldsymbol{k}_1)\rangle&=A_3(k_1,k_2,k_3)\,\Pi_{\mu\nu}^{\alpha\beta}( \boldsymbol{k}_3)\,k_2^\alpha\,k_2^\beta
%\end{align}
%%
A simple counting of the number of unknowns and independent equations makes it clear that the higher spin equation is not powerful enough to solve for all the form factors individually. Solving for two of the form factors  $A_2(k_1,k_2,k_3)$ and $A_3(k_1,k_2,k_3)$ that appear in $\langle J_0( \boldsymbol{k}_1)\,T_{\mu\nu}( \boldsymbol{k}_2)\,J_0(\boldsymbol{k}_3)\rangle$ and  $\langle J_0( \boldsymbol{k}_1)\,J_0( \boldsymbol{k}_2)\,T_{\mu\nu}( \boldsymbol{k}_3)\rangle$ respectively in terms of the form factor in $\langle T_{\mu\nu}( \boldsymbol{k}_1)\,J_0( \boldsymbol{k}_2)\,J_0( \boldsymbol{k}_3)\rangle$, viz. $A_1(k_1,k_2,k_3)$ we obtain :
\begin{align}
\label{A2A3intermsofA1}
A_2(A_1)&=\frac{k_1^3-k_1^2(4A_1k_2^2+k_3)+k_2^2(k_2-k_3)(4A_1(k_2+k_3)-1)}{4k_1^2(k_1^2-k_2^2-k_3^2)}\cr
A_3(A_1)&=\frac{k_1^3-k_1^2(4A_1k_3^2+k_2)-k_3^2(k_2-k_3)(4A_1(k_2+k_3)-1)}{4k_1^2(k_1^2-k_2^2-k_3^2)}\,.
\end{align}
One can easily verify that the form factors obtained by an explicit computation in \eqref{formfactorTOO} satisfy the above relations. However,
to completely determine the form factors we require one more constraint and that is provided by imposing conformal invariance. It can be easily checked that given $A_1$ satisfies the dilatation Ward identity, $A_{2}(A_1)$ and $A_{3}(A_1)$ in \eqref{A2A3intermsofA1} satisfy the identity trivially, i.e. it does not give us any new constraint. Therefore to solve for the form factors, we 
combine the higher spin equation with the primary Ward identities.  After a series of steps that involve imposing the primary Ward identity one arrives at a first order differential equation of the following kind :
\begin{align}
\label{FODEmain}
g_1(k_1,k_2,k_3)+\widetilde g_2(k_1,k_2,k_3)A_1+\widetilde g_4(k_1,k_2,k_3)\frac{\partial A_1}{\partial k_2}=0\,.
\end{align}
For details of this computation and the explicit form of $g_1(k_1,k_2,k_3)$, $\widetilde g_2(k_1,k_2,k_3)$ and $\widetilde g_4(k_1,k_2,k_3)$ see Appendix \ref{HSEQ4fermion}.
Since this is a first order differential equation this can in principle be solved. Here we simply note that the form factor $A_1(k_1,k_2,k_3)$ obtained in \eqref{formfactorTOO} satisfies this first order differential equation. Thus the three-point correlation function $\langle T_{\mu\nu}\,J_0\,J_0\rangle$ in the fermionic theory is determined by the higher spin equations aided by conformal invariance.

\section{Discussion}

In this paper we explicitly computed various parity odd  as well as parity even  correlation functions involving scalar and spinning operators in three dimensional free theories. In particular we developed a basis for the transverse part of parity odd correlators. We demonstrated that the techniques involving Schouten identity and the inversion of momenta provide an efficient way to compute correlators. We then explored the higher spin equations for free theories and demonstrated that some of the three-point functions could be solved for using only the higher spin equations. However for four-point functions we could only verify that our explicit results solve the higher spin equation and not actually solve for them.

There are a few immediate generalisations of our work. Generalising our results to interacting theories such as the Chern-Simons matter theories is interesting \cite{wip}. One may use slightly broken higher spin symmetry to solve for correlators in those theories. It would also be exciting to find out using higher spin equations structures which are not dictated by free theories. Another interesting question to ask is if higher spin equations can be used to understand contact terms which appear in momentum space quite often. 

It would also be interesting to explore the double copy structure of CFT correlators \cite{Farrow_2019,wip1}. Another interesting direction to pursue is to understand the spin and weight raising operators of \cite{arkanihamed2018cosmological,baumann2019cosmological,baumann2020cosmological} for parity odd correlators. 
  
One can also try to compute correlation functions in momentum space as invariants of higher spin symmetry, following the position space analysis in \cite{Didenko:2012tv,Didenko:2013bj}.
It would also be interesting to examine higher spin equations in the context of \cite{gillioz2020scattering}. In the case of four-point correlators, we saw that although explicit results solved the higher spin equation, one could not solve the equations to obtain the correlators. This was because there were more variables to solve for than equations. The large number of variables stemmed from a lack of understanding of conformal cross-ratios in momentum space. We believe that the recent work \cite{gillioz2020scattering} gives a way to deal with this issue. We hope to address these issues in the near future.
 %If we assume that the formalism in \cite{gillioz2020scattering} can be extended to that of spinning operators, then the higher spin equation can be cast in terms of such form factors. This would then amount to solving for fewer variables.
%%%
%\section*{Appendix}
%

\section*{Acknowledgments}
We would like to thank A.~Mehta for extensive discussions. We acknowledge T.~Sharma, D.~Ghosh, A.~Nizami, S.~Mishra, S.~Sharma for fruitful discussions. RRJ thanks IISER Pune for hospitality where most of the work was done. Research of SJ and VM is supported by the Ramanujan Fellowship. The work of RRJ is supported by the MIUR PRIN Contract 2015 MP2CX4 ``Non-perturbative Aspects Of Gauge Theories And Strings''. The work of RRJ is also partially supported by ``Fondi Ricerca Locale dell’Universit\`a del Piemonte Orientale''. SJ and VM would like to acknowledge their debt to the people of India for their steady support of research in basic sciences. 
\appendix
	\section{Compendium of Integrals}\label{Comp}
	In this section we provide a compendium of integrals which we have used for the computation of the various correlators involving scalar, spin-1 current and stress-tensor operators. We give a brief derivation of these in this Appendix. Note that in our notation $p_{ij}^{(m)}=|\boldsymbol{p}_{i}-\boldsymbol{p}_{j}|$ whereas $p_{ij}=|\boldsymbol{p}_{i}+\boldsymbol{p}_{j}|$.  We also use the following notation :
	\begin{align}
	\int_l=\int \frac{d^{3} l}{(2 \pi)^{3}}\,.
	\end{align} 
	Let us note that we are using dimensional regularization. If one uses PV reduction to evaluate integrals with four denominators, it soon becomes extremely cumbersome and inefficient. In this Appendix we use the inversion method combined with the fact that in three dimensions any momentum can be decomposed in terms of three independent momentum, to compute one loop integrals efficiently.  
	\subsection{Scalar Integrals}\label{ScInts}
	Below is the list of scalar integrals that we encountered in our computations :
	\begin{align}
	\int_l \frac{1}{(l^2)(l+p)^2}&=\frac{1}{8 p}\label{SC2pt1}\\
	\int_l \frac{1}{l^{2}\left(l+p_{1}\right)^{2}\left(l+p_{2}\right)^{2}}&=\frac{1}{8\,p_{1}\,p_2\,p_{12}^{(m)}}\label{SC3pt1}\\
	\int_l\frac{1}{l^{2}\left(l+p_{1}\right)^{2}\left(l+p_{2}\right)^{2}\left(l+p_{3}\right)^{2}}&=\frac{1}{8 p_1\,p_2\,p_3}\frac{p_1\,p_2\,p_{12}^{(m)}+p_2\,p_3\,p_{23}^{(m)}+p_3\,p_1\,p_{13}^{(m)}+p_{12}^{(m)}\,p_{13}^{(m)}\,p_{23}^{(m)}}{p_{12}^{(m)}\,p_{13}^{(m)}\,p_{23}^{(m)} (p_1\,p_{23}^{(m)}+p_2\,p_{13}^{(m)}+ p_3\,p_{12}^{(m)})}\,.\label{SC4pt1}
	\end{align}
	For the box scalar integration \eqref{SC4pt1}, we used the inversion technique {\cite{Anninos:2017eib,Yacoby:2018yvy}}. This technique involves inverting all the momenta including the integration variable as follows $l^{\mu}=\frac{\tilde{l}^{\mu}}{\tilde{l}^2}$ and $p_i=\frac{\tilde{P}_i}{\tilde{P}^2}$ ($i=1,2,3$). The integration measure changes as $\int d^3l=\int \frac{d^3\tilde{l}}{\tilde{l}^2}$. Following this inversion, one observes that the integral reduces to simpler known integrals in the inverted variables. We then use the results for known integrals and invert back momenta to obtain the result in terms of the original momenta.
	\subsection{Vector Integrals}
	The list of vector integrals is as follows :
	\begin{align}
	&\int_l\frac{l^{\mu}}{l^2(l+p_1)^2}=-\frac{p_{1}^{\mu}}{16 p_1},~~
	\int_l\frac{l^{\mu}}{l^2(l+p_1)^2(l+p_2)^2}=-\frac{p_2 p_1^{\mu}  + p_1 p_2^\mu}{8 p_1  p_2  p_{12}^{(m)}(p_1 + p_2+ p_{12}^{(m)} )}\label{Vec3pt1}
	\end{align}
	\begin{align}
	&\int_l\frac{l^{\mu}}{l^{2}\left(l+p_{1}\right)^{2}\left(l+p_{2}\right)^{2}\left(l+p_{3}\right)^{2}}\notag\\&~~~~=-\frac{1}{8(p_1 p_{23}^{(m)}+p_2 p_{13}^{(m)} +p_3 p_{12}^{(m)} )}\bigg[\frac{p_1^{\mu}}{ p_1 p_{12}^{(m)} p_{13}^{(m)} }+\frac{p_2^{\mu}}{p_2 p_{12}^{(m)}  p_{23}^{(m)}}+\frac{p_3^{\mu}}{p_3 p_{13}^{(m)} p_{23}^{(m)} }\bigg]\,.\label{Vec4pt1}
	\end{align}
	To evaluate the integral in \eqref{Vec4pt1} one can use the inversion technique again. 
	\subsection{Two Tensor Integrals}\label{2tIntegral}
	The list of two index tensor integrals that we encountered are the following :
	\begin{align}
	\label{2T3pt1}
	&\int_l \frac{l^{\mu}l^{\nu}}{l^2(l+p_1)^2(l-p_2)^2}\cr
	&\hspace{1cm}=g^{\mu\nu}\,A(p_1,p_2,p_{12})-(p_1^{\mu}\,p_2^{\nu}+p_2^{\mu}\,p_1^{\nu})\,B(p_1,p_2,p_{12})
	+p_1^{\mu}\,p_1^{\nu}\,C(p_1,p_2,p_{12})+p_2^{\mu}\,p_2^{\nu}\,C(p_2,p_1,p_{12})\cr
&\text{where} \quad A(p_1,p_2,p_{12})=\frac{1}{16 ~(p_1+p_2+p_{12})}\cr
&\text{          } \quad\quad\,\,\,\,\,\,\,B(p_1,p_2,p_{12})=\frac{1}{16~p_{12}(p_1+p_2+p_{12})^2}\cr
&\text{          } \quad\quad\,\,\,\,\,\,\,C(p_1,p_2,p_{12})=\frac{(p_1+2p_2+p_{12})}{16~ p_1 p_{12}(p_1+p_2+p_{12})^2}\,.
	\end{align}
The evaluation of \eqref{2T3pt1} is straightforward using the standard technique of PV reduction and we do not describe it here.
	However, two tensor integrals of the following kind with four denominators : 
	\begin{align}
\int_l\frac{l^{\mu}l^{\nu}}{l^{2}\left(l+p_{1}\right)^{2}\left(l+p_{2}\right)^{2}\left(l+p_{3}\right)^{2}}\label{2TInt4}
	\end{align}
	become very cumbersome using PV reduction.  Here we use a method analogous to the scalar integration in \cite{vanNeerven:1983vr} utilising the following Schouten identity :
\begin{align}\label{lmu}
l^{\mu} =\frac{1}{\epsilon^{p_1 p_2 p_3}}\bigg[\epsilon^{\mu p_2 p_3} l\cdot p_1+\epsilon^{p_1\mu  p_3}l\cdot p_2+\epsilon^{p_1 p_2\mu  }l\cdot p_3\bigg]
\end{align}
where $\epsilon^{\mu p_i p_j}$ and $\epsilon^{ p_1 p_2p_3}$ are defined following \eqref{enot}.
%$\epsilon^{\mu p_i p_j}=\epsilon^{\mu \alpha \beta}p_{i\alpha}p_{j\beta}$ and
%$\epsilon^{ p_1 p_2p_3}=\epsilon^{\mu \alpha \beta}p_{1\mu}p_{2\alpha}p_{3\beta}$.
%\begin{align}
%\epsilon^{\mu p_i p_j}&=\epsilon^{\mu \alpha \beta}p_{i\alpha}p_{j\beta}\hspace{1cm}
%\epsilon^{ p_1 p_2p_3}=\epsilon^{\mu \alpha \beta}p_{1\mu}p_{2\alpha}p_{3\beta}
%\end{align}
We substitute the Schouten identity for $l^{\mu}$ in the required integral \eqref{2TInt4} to obtain :
\begin{align}
\int_l\frac{l^{\mu}l^{\nu}}{l^{2}\left(l+p_{1}\right)^{2}\left(l+p_{2}\right)^{2}\left(l+p_{3}\right)^{2}}=\int_l\frac{1}{\epsilon^{p_1 p_2 p_3}}\frac{l^{\nu} (\epsilon^{\mu p_2 p_3}\,l\cdot p_1+\epsilon^{p_1\mu  p_3}\,l\cdot p_2+\epsilon^{p_1 p_2\mu  }\,l\cdot p_3)}{l^{2}\left(l+p_{1}\right)^{2}\left(l+p_{2}\right)^{2}\left(l+p_{3}\right)^{2}}\,.
\end{align}
Note that the Schouten identity shifts the $\mu$ index from $l$ to $\epsilon$.
We can now use  $l\cdot p_i=\frac{1}{2}((l+p_i)^2-l^2-p_i^2)$ for $i=1,2,3$ to reduce the above integral to the integrals we described in \eqref{SC4pt1} and \eqref{Vec3pt1}. This method can be generalised to integrals with more complicated tensor structure. Performing what we described above we obtain :
	\begin{align}
	\int_l&\frac{l^{\mu}l^{\nu}}{l^{2}\left(l+p_{1}\right)^{2}\left(l+p_{2}\right)^{2}\left(l+p_{3}\right)^{2}}\notag\\=&\frac{1}{\epsilon^{p_1 p_2 p_3}}\epsilon^{\mu p_2 p_3}\bigg[A_1(p_1,p_2,p_3) p_1^{\nu}+B_1(p_1,p_2,p_3) p_2^{\nu}+C_1(p_1,p_2,p_3) p_3^{\nu}\bigg]\notag\\
	&+\frac{1}{\epsilon^{p_1 p_2 p_3}}\epsilon^{p_1\mu  p_3}\bigg[A_2(p_1,p_2,p_3) p_1^{\nu}+B_2(p_1,p_2,p_3) p_2^{\nu}+C_2(p_1,p_2,p_3) p_3^{\nu}\bigg]\notag\\
	&+\frac{1}{\epsilon^{p_1 p_2 p_3}}\epsilon^{\mu p_2 p_3}\bigg[A_3 (p_1,p_2,p_3)p_1^{\nu}+B_3(p_1,p_2,p_3) p_2^{\nu}+C_3(p_1,p_2,p_3) p_3^{\nu}\bigg]\label{2T4pt1}
	\end{align}
	where $A_i,B_i,C_i$ for $i=1,2,3$ are as follows :
\begin{align}\label{2tnsrintR1}
	A_1(p_1,p_2,p_3) &= \frac{p_1 (p_{12}^{(m)}+p_{13}^{(m)}+2 p_{23}^{(m)})+p_{12}^{(m)} p_3+p_{13}^{(m)} p_2}{16 p_{12}^{(m)} p_{13}^{(m)} (p_{12}^{(m)}+p_{13}^{(m)}+p_{23}^{(m)}) (p_1 p_{23}^{(m)}+p_{12}^{(m)} p_3+p_{13}^{(m)} p_2)}\nonumber\\[10pt]
	B_1(p_1,p_2,p_3) &=\frac{1}{16 p_{23}^{(m)}}\bigg[\frac{p_1^2}{p_2 p_{12}^{(m)} (p_1 p_{23}^{(m)}+p_{12}^{(m)} p_3+p_{13}^{(m)}p_2)}+\frac{1}{p_{12}^{(m)} (p_{12}^{(m)}+p_{13}^{(m)}+p_{23}^{(m)})}\nonumber\\[5pt]
	&\hspace{2cm}-\frac{1}{p_2  (p_2+p_3+p_{23}^{(m)})}\bigg]\nonumber\\[10pt]
	C_1(p_1,p_2,p_3)& =\frac{1}{16 p_{23}^{(m)}}\bigg[\frac{p_1^2}{ p_{13}^{(m)}  p_3 (p_1 p_{23}^{(m)}+p_{12}^{(m)} p_3+p_{13}^{(m)} p_2)}+\frac{1}{ p_{13}^{(m)}  (p_{12}^{(m)}+p_{13}^{(m)}+p_{23}^{(m)})}\nonumber\\[5pt]
	&\hspace{2cm}-\frac{1}{ p_3 (p_2+p_{23}^{(m)}+p_3)}\bigg]\nonumber\\[10pt]
	A_2(p_1,p_2,p_3)&=\frac{1}{16 p_{13}^{(m)}}\bigg[\frac{p_2^2}{p_1 p_{12}^{(m)}  (p_1 p_{23}^{(m)}+p_{12}^{(m)} p_3+p_{13}^{(m)} p_2)}-\frac{1}{ p_1 (p_1+p_{13}^{(m)}+p_3)}\nonumber\\[5pt]
	&\hspace{2cm}+\frac{1}{p_{12}^{(m)}  (p_{12}^{(m)}+p_{13}^{(m)}+p_{23}^{(m)})}\bigg]\nonumber\\[10pt]
	B_2(p_1,p_2,p_3)&=
	\frac{p_{23}^{(m)} (p_1+p_2)+p_{12}^{(m)} (p_2+p_3)+2 p_{13}^{(m)} p_2}{16 p_{12}^{(m)} p_{23}^{(m)} (p_{12}^{(m)}+p_{13}^{(m)}+p_{23}^{(m)}) (p_1 p_{23}^{(m)}+p_{12}^{(m)} p_3+p_{13}^{(m)} p_2)}\nonumber\\[10pt]
%%%	
C_2(p_1,p_2,p_3)&=\frac{1}{16 p_{13}^{(m)}}\bigg[\frac{p_2^2}{ p_{23}^{(m)}p_3 (p_1 p_{23}^{(m)}+p_{12}^{(m)} p_3+p_{13}^{(m)} p_2)}-\frac{1}{p_3 (p_1+p_{13}^{(m)}+p_3)}\nonumber\\[5pt]
&\hspace{2cm}+\frac{1}{ p_{23}^{(m)}(p_{12}^{(m)}+p_{13}^{(m)}+p_{23}^{(m)})}\bigg]\nonumber\\[10pt]
A_3(p_1,p_2,p_3)&=\frac{1}{16p_{12}^{(m)}}\bigg[\frac{p_3^2}{p_1  p_{13}^{(m)} (p_1 p_{23}^{(m)}+p_{12}^{(m)} p_3+p_{13}^{(m)} p_2)}-\frac{1}{p_1  (p_1+p_{12}^{(m)}+p_2)}\nonumber\\[5pt]
&\hspace{2cm}+\frac{1}{p_{13}^{(m)} (p_{12}^{(m)}+p_{13}^{(m)}+p_{23}^{(m)})}\bigg]\nonumber\\[10pt]
B_3(p_1,p_2,p_3)&=\frac{1}{16 p_{12}^{(m)} }\bigg[\frac{p_3^2}{p_2 p_{23}^{(m)} (p_1 p_{23}^{(m)}+p_{12}^{(m)} p_3+p_{13}^{(m)} p_2)}-\frac{1}{ p_2 (p_1+p_{12}^{(m)}+p_2)}\nonumber\\[5pt]
&\hspace{2cm}+\frac{1}{ p_{23}^{(m)} (p_{12}^{(m)}+p_{13}^{(m)}+p_{23}^{(m)})}\bigg]\nonumber\\[10pt]
C_3(p_1,p_2,p_3)&=\frac{p_1p_{23}^{(m)}+p_3(2 p_{12}^{(m)}+p_{23}^{(m)})+p_{13}^{(m)} (p_2+p_3)}{16 p_{13}^{(m)} p_{23}^{(m)} (p_{12}^{(m)}+p_{13}^{(m)}+p_{23}^{(m)}) (p_1p_{23}^{(m)}+p_{12}^{(m)} p_3+p_{13}^{(m)} p_2)}\,.
	\end{align}
	\section{Wick Contractions and other details}
\label{Wick contractions}
\subsection{$\langle T_{\mu\nu}\,J_0\,J_0\,J_0 \rangle$}\label{WCJ24}
Here we give the computational details of the correlator $\langle T_{\mu\nu}J_0 J_0 J_0 \rangle$ in the free bosonic theory. The stress-tensor in momentum space as given in Section \ref{Theory} is :
\begin{align}
T_{\mu\nu}(k)&=\int d^3l\bigg[-\frac{3}{8}(l_{\mu}(k-l)_{\nu}+(k-l)_{\mu}l_{\nu})+\frac{1}{8}((k-l)_{\mu}(k-l)_{\nu}+l_{\mu}l_{\nu})\notag\\&\hspace{1.5cm}+\frac{1}{4}g_{\mu \nu}\,l\cdot (k-l)-\frac{g_{\mu\nu}}{24}(l^2+(k-l)^2)\bigg] \bar{\phi}(l) \phi(k-l)\notag\\
&=\int d^3l\,H_{\mu \nu}(l,k)\,\bar{\phi}(l) \phi(k-l)\,.
\end{align}
The Wick contraction $1\bar{2}~2\bar{3}~3\bar{4}~4\bar{1}$ and its complex conjugate $1\bar{4}~4\bar{3}~3\bar{2}~2\bar{1}$ are equal and lead to the following :
\begin{align}
G^{\mu\nu}_{1}&=\int_{l_1l_2l_3l_4} \langle\bar{\phi}(l_1) H^{\mu\nu}\phi(k_1-l_1)\bar{\phi}(l_2)\phi(k_2-l_2)\bar{\phi}(l_3)\phi(k_3-l_3)\bar{\phi}(l_4)\phi(k_4-l_4)\rangle\notag\\
=&\int_{l_1l_2l_3l_4} \frac{1}{4}\frac{ (k_1^{\mu} k_1^{\nu} - 4 k_1^{\nu} l_1^{\mu} - 4 k_1^{\mu} l_1^{\nu} + 8 l^{\mu} l^{\nu})}{l_1^2 l_2^2 l_3^2 l_4^2}
+g^{\mu \nu}\bigg[\frac{1}{2}\int\frac{ \left(l_1.k_1-l_1^2\right)}{l_1^2 l_2^2 l_3^2 l_4^2}-\frac{1}{12}\int  \frac{\left(l^2+(l-k)^2\right)}{l_1^2 l_2^2 l_3^2 l_4^2}\bigg]\notag\\&\delta(k_1-l_1+l_2)\delta((k_2-l_2+l_3)\delta(k_3-l_3+l_4)\delta(k_4-l_4+l_1)
\end{align}
We use the delta functions to set :
\begin{align}
l_1&=l_2+k_1\cr
l_4&=l_2-k_2-k_3\cr
l_3&=l_2-k_2
\end{align}
This leads to :
\begin{align}
G^{\mu\nu}_{1}&=\int d^3l\frac{1}{4} \frac{(k_1^{\mu} k_1^{\nu} + 4 k_1^{\nu} l^{\mu} +4 k_1^{\mu} l^{\nu} + 8 l^{\mu} l^{\nu}))}{l^2(l+k_1)^2(l-k_2)^2(l-k_2-k_3)^2}\nonumber\\[5pt]&+g^{\mu \nu}\bigg[-\frac{1}{2}\int d^3l\frac{ l^2 + l.k_1}{l^2(l+k_1)^2(l-k_2)^2(l-k_2-k_3)^2}-\frac{1}{12}\int d^3l  \frac{(l+k_1)^2+l^2}{l^2(l+k_1)^2(l-k_2)^2(l-k_2-k_3)^2}\bigg]
\end{align}
Similarly, the Wick contractions $1\bar{3}~~3\bar{4}~4\bar{2}~~2\bar{1}$ and  $1\bar{2}~2\bar{4}~4\bar{3}~3\bar{1}$ lead to :
\begin{align}
G^{\mu\nu}_{2}&=\frac{1}{4}\int d^3l \frac{(k_1^{\mu} k_1^{\nu} - 4 k_1^{\nu} l^{\mu} -4 k_1^{\mu} l^{\nu} + 8 l^{\mu} l^{\nu}))}{l^2(l-k_1)^2(l+k_2)^2(l-k_1-k_3)^2}\nonumber\\[5pt]&+g^{\mu \nu}\bigg[-\frac{1}{2}\int d^3l\frac{ l^2 - l.k_1}{l^2(l-k_1)^2(l+k_2)^2(l-k_1-k_3)^2}-\frac{1}{12}\int d^3l  \frac{(l-k_1)^2+l^2}{l^2(l-k_1)^2(l+k_2)^2(l-k_1-k_3)^2}\bigg]
\end{align}
The Wick Contractions $1\bar{3}~~3\bar{2}~2\bar{4}~~4\bar{1}$ and $1\bar{4}~4\bar{2}~2\bar{3}~3\bar{1}$ lead to :
\begin{align}
G^{\mu\nu}_{3}&=\frac{1}{4}\int d^3l \frac{(k_1^{\mu} k_1^{\nu} + 4 k_1^{\nu} l^{\mu} +4 k_1^{\mu} l^{\nu} + 8 l^{\mu} l^{\nu}))}{l^2(l+k_1)^2(l-k_3)^2(l-k_2-k_3)^2}\nonumber\\[5pt]&+g^{\mu \nu}\bigg[-\frac{1}{2}\int d^3l\frac{ l^2 + l.k_1}{l^2(l+k_1)^2(l-k_3)^2(l-k_2-k_3)^2}-\frac{1}{12}\int d^3l  \frac{(l+k_1)^2+l^2}{l^2(l+k_1)^2(l-k_3)^2(l-k_2-k_3)^2}\bigg]
\end{align}
The full correlator is therefore given by the following sum :
\begin{align}
\langle T^{\mu\nu}( \boldsymbol{k}_1) J_{0}( \boldsymbol{k}_2) J_0( \boldsymbol{k}_3)J_0( \boldsymbol{k}_4)\rangle=G^{\mu\nu}_{1}+G^{\mu\nu}_{2}+G^{\mu\nu}_{3}\,.
\end{align}
The evaluation of the above integrals is tedious because of the presence of integrals of the form \eqref{2T4pt1}. However, one may use a clever trick to perform the above integral and obtain the result.  The trick is to separate the $G^{\mu\nu}_{i}$ as follows :
\begin{align}
G^{\mu\nu}_{1}=\frac{a\,\Pi^{\mu\nu}_{\alpha\beta}(\boldsymbol{k}_1)\,l^{\alpha}\,l^{\beta}}{l^2(l+k_1)^2(l-k_2)^2(l-k_2-k_3)^2}+X^{\mu\nu}\,,
\end{align}
where $\Pi^{\mu\nu}_{\alpha\beta}$ is the projector defined in \eqref{projk1k1}. $G^{\mu\nu}_{2}$  and $G^{\mu\nu}_{3}$ may also be separated as above. Here $a$ is a constant which is chosen such that no $l^{\mu}l^{\nu}$ term appears in $X^{\mu\nu}$. This way of separating the integral makes the calculation more efficient as we do not require the complete integral \eqref{2T4pt1} but only a projected part of it.
The computation of the same correlator in the fermionic theory is quite similar and we do not provide the details here.

\subsection{$\langle J_0\,J_0\,J_0\,J_0\,J_0\rangle$}\label{j0B5}

Here we give details of the computation of the integral required to compute the five-point function of the scalar operator in free bosonic theory. The method of inversion of momenta for this computation is a bit tedious. Hence we employ the method of determining the integral required to compute the five-point function in terms of integrals appearing in the  four-point function through an interesting identity known as the Schouten identity as suggested in \cite{vanNeerven:1983vr}. The Schouten identity for a three dimensional vector is as follows :
\begin{align}
\epsilon^{p_1p_2p_3} l^{\mu}=\epsilon^{\mu p_2 p_3} l\cdot p_1+\epsilon^{p_1\mu  p_3}l\cdot p_2+\epsilon^{p_1 p_2\mu  }l\cdot p_3\,.\label{Sch}
\end{align}	
As explained in \cite{vanNeerven:1983vr}, the presence of another independent momentum $p_4$ simplifies the computation of the integral\footnote{One can also perform a similar computation to determine the integral that appears in the computation of four-point from the  integrals for three-point function. However, in this case the method of inversion that we utilised is more efficient and simple. See \cite{vanNeerven:1983vr} for details.} as one may construct a linear equation by dotting the above identity with the momentum $p_4$. This leads to the following :
\begin{align}
\int_l\frac{\epsilon^{p_1p_2p_3} l\cdot p_4-(\epsilon^{p_4 p_2 p_3} l\cdot p_1+\epsilon^{p_1p_4 p_3}l\cdot p_2+\epsilon^{p_1 p_2 p_4  }l\cdot p_3)}{l^2(l+p_1)^2(l+p_2)^2(l+p_3)^2(l+p_4)^2}=0\,. \label{Inteq9}
\end{align}
We may now express $l\cdot p_i=\frac{1}{2}((l+p_i)^2-l^2-p_i^2)$ and reduce most of the above integrals to integrals that appear in the computation of the four-point function. This leads to an interesting relation which can be expressed in the following compact notation :
\begin{align}
F(p_1,p_2,p_3,p_4,p_5)&=\int_l \frac{1}{l^2(l+p_1)^2(l+p_2)^2(l+p_3)^2(l+p_4)^2}\notag\\&=\frac{1}{f }\bigg[\epsilon^{p_1p_2p_3}  E_{0123}-\epsilon^{p_4 p_2 p_3} E_{0234}-\epsilon^{p_1 p_4 p_3} E_{0134}-\epsilon^{p_1 p_2 p_4} E_{0124}\notag\\&\hspace{1cm}+(\epsilon^{p_4 p_2 p_3}+\epsilon^{p_1 p_4 p_3}+\epsilon^{p_1 p_2 p_4} -\epsilon^{p_1p_2p_3} )E_{1234}\bigg]
\notag\\&=\frac{1}{f }\bigg[\epsilon^{p_1p_2p_3}(E_{0123}-E_{1234})-\epsilon^{p_2p_3p_4}(E_{0234}-E_{1234})\notag\\&\hspace{1cm}+\epsilon^{p_3p_4p_1}( E_{0134}-E_{1234})-\epsilon^{p_4p_1p_2}( E_{0124}-E_{1234})\bigg]\,,\label{E5}
\end{align}
where
\begin{align}
f&=p_4^2\epsilon^{p_1p_2p_3}-(p_1^2\epsilon^{p_4p_2p_3}+p_2^2 \epsilon^{p_1p_4p_3}+p_3^2\epsilon^{p_1p_2p_4})\nonumber\\[5pt]
&=-p_1^2\epsilon^{p_2p_3p_4}+p_2^2 \epsilon^{p_3p_4p_1}-p_3^2\epsilon^{p_4p_1p_2}+p_4^2\epsilon^{p_1p_2p_3}\cr
E_{0ijk}&=\int_l \frac{1}{l^2(l+p_i)^2(l+p_j)^2(l+p_k)^2}\notag\\&=\frac{1}{8 p_i p_j p_k }\frac{p_i  p_j p_{ij}^{(m)}+p_j p_k p_{jk}^{(m)}+p_k p_i p_{ik}^{(m)}+p_{ij}^{(m)} p_{ik}^{(m)}p_{jk}^{(m)}}{ p_{ij}^{(m)}p_{ik}^{(m)}   p_{jk}^{(m)}(p_i p_{jk}^{(m)}+ p_jp_{ik}^{(m)}+ p_k p_{ij}^{(m)}  )}\notag\\
E_{1234}&=\int_l \frac{1}{(l+p_1)^2(l+p_2)^2(l+p_3)^2(l+p_4)^2}\notag\\&=\int_l \frac{1}{l^2(l+(p_2-p_1))^2(l+(p_3-p_1))^2(l+(p_4-p_1))^2}\notag\\&=
E_{0p_{21}^{(m)}p_{31}^{(m)}p_{41}^{(m)}}\,.
\end{align}
Let us now consider the integral we have for the four-point function :
\begin{align}
H(k_{1}, k_{2}, k_{3}, k_{4}, k_{5})=\int_l \frac{d^{3} l}{(2 \pi)^{3}} \frac{1}{l^{2}(l+k_{1})^{2}(l+k_{1}+k_{2})^{2}(l+k_{1}+k_{2}+k_3)^{2}(l-k_5)^2}\,.
\end{align}
Utilising the identity in \eqref{E5} with the identifications $p_1=k_1,p_2=k_{12},p_3=k_{123},p_4=-k_{5}=k_{1234}$ we obtain :
\begin{align}
H(k_{1}, k_{2}, k_{3}, k_{4}, k_{5})=F(k_1,k_{12},k_{123},k_{1234})
\end{align}
where $F(k_1,k_{12},k_{123},k_{1234})$ can be obtained from \eqref{E5} by the integrals that appear in the four-point correlator.
%%%%
%%%
\section{Details of the higher spin equation}
\label{HSEQ4fermion}
The primary Ward identity on the form factors is given by :
\begin{align}
\label{PWI}
K_{ij}A_{1,2,3}=0,\quad i,j=1,2,3\,,
\end{align}
where $K_{ij}=K_i-K_j$ and 
\begin{align}
K_i=\frac{\partial^2}{\partial k_i^2}+\frac{4-2\Delta_i}{k_i}\frac{\partial}{\partial k_i}\,.
\end{align}
Given $A_2$ as a function of $A_1$ \eqref{A2A3intermsofA1} we impose \eqref{PWI} by acting $K_{12}$ on $A_2(A_1)$ :
\begin{align}
\label{PWIA2}
\frac{\partial^2 A_2(A_1)}{\partial k_2^2}-\frac{\partial^2 A_2(A_1)}{\partial k_1^2}-\frac{2}{k_2}\frac{\partial A_2(A_1)}{\partial k_1}=0\,.
\end{align}
We now impose in \eqref{PWIA2} the following primary Ward identity that $A_1$ must satisfy :
\begin{align}
\frac{\partial^2 A_1}{\partial k_2^2}-\frac{\partial^2 A_1}{\partial k_1^2}-\frac{2}{k_2}\frac{\partial A_1}{\partial k_1}=0\,.
\end{align}
This reduces the second order differential equation to a first order differential equation :
\begin{align}
\label{FODE}
g_1(k_1,k_2,k_3)+g_2(k_1,k_2,k_3)A_1+g_3(k_1,k_2,k_3)\frac{\partial A_1}{\partial k_1}+g_4(k_1,k_2,k_3)\frac{\partial A_1}{\partial k_2}=0\,,
\end{align}
where 
\begin{align}
g_1&=\frac{1}{2k_1^4(k_1^2-k_2^2-k_3^2)^3}\big(-2k_1^7+3k_1^6k_3+2k_1^5(k_2^2-k_3^2)+3k_2^2(k_2-k_3)(k_2^2+k_3^2)^2\cr
&\hspace{3cm}+k_1^4(5k_2^3-9k_2^2k_3+2k_3^3)-k_1^2(8k_2^5-9k_2^4k_3+4k_2^3 k_3^2-4k_2^2k_3^3+k_3^5)\big)\cr
g_2&=\frac{2}{k_1^4(k_1^2-k_2^2-k_3^2)^3}\big(k_1^8-k_1^6k_3^2-3k_2^2(k_2^2-k_3^2)(k_2^2+k_3^2)^2-k_1^4(6k_2^4-7k_2^2k_3^2+k_3^4)\cr
&\hspace{4cm}+k_1^2(8k_2^6-3k_2^4k_3^2-2k_2^2k_3^4+k_3^6)\big)\cr
g_3&=-\frac{2k_2^2(k_1^4+k_2^4-k_3^4-2k_1^2(k_2^2-2k_3^2))}{k_1^3(k_1^2-k_2^2-k_3^2)^2}\cr
g_4&=-\frac{2k_2(k_1^4-2k_1^2k_2^2+k_2^4+4k_2^2k_3^2-k_3^4}{k_1^2(k_1^2-k_2^2-k_3^2)^2}
\end{align}
We make use of the dilatation Ward identity that $A_1(k_1,k_2,k_3)$ satisfies :
\begin{align}
\left(\sum_{i=1}^3\,k_i\frac{\partial}{\partial k_i}+1\right)A_1(k_1,k_2,k_3)=0
\end{align}
and the symmetry of the correlator under $k_2\leftrightarrow k_3$ exchange 
to obtain \eqref{FODEmain}
where 
\begin{align}
\widetilde g_2(k_1,k_2,k_3)&=\left[g_2(k_1,k_2,k_3)-\frac{1}{k_1}g_3(k_1,k_2,k_3)\right]\cr
\widetilde g_4(k_1,k_2,k_3)&=g_4(k_1,k_2,k_3)-g_3(k_1,k_2,k_3)\frac{(k_2+k_3)}{k_1}\,.
\end{align}

%\bibliography{FF}

\bibliographystyle{JHEP}

\end{document}